# AI-guided high-throughput discovery of iridium- and ruthenium-free palladium-oxide catalysts for durable acidic oxygen evolution

Ken J. Jenewein[a,1], Faezeh Habib Zadeh[a,1], Xiaoxiao Wang[a], Gustavo Malkomes[a], Huafan Zhang[a], Natalie Page[a], Jae Jin Bang[a], Peter J. Santiago[a], Karla V. Contreras[a], Katherine K. Li[a], Allison Perna[a], Lorena M. Britton[a], Fahrettin Kilic[a], Kevin J. Cruse[a], Armin Taheri[a], Krishnanand Mallayya[a], Harley Quinn[a], Rebecca A. Durr[a], Peter A. Beaucage[a], Santiago Miret[a], John M. Gregoire[a,2], Rafael Gómez-Bombarelli[a,2]

[a]Lila Sciences, Inc., Cambridge, MA 02141, USA

[1]K.J.J. and F.H.Z. contributed equally to this work

[2]To whom correspondence may be addressed. Email: jgregoire@lila.ai and rgbombarelli@lila.ai

**Author Contributions:** J.M.G. and R.G.B. conceived and supervised the project, with R.G.B. leading the AI/ML strategy and J.M.G. leading the autonomous-science strategy; K.J.J., F.H.Z., X.W., G.M., H.Z., N.P., J.M.G., and R.G.B. designed the research; K.J.J. and F.H.Z. led experiments and the electrochemical workflow; K.J.J., F.H.Z., P.J.S., K.V.C., and K.K.L. executed electrochemical workflows; N.P. led catalyst synthesis; N.P. and F.K. executed synthesis workflows; H.Z. led materials characterization; H.Z., L.M.B., and A.P. executed characterization workflows; J.J.B. contributed to characterization data discussions; G.M. and X.W. led the ML development and executed computations; K.J.C. contributed to ML development; A.T. and K.M. contributed to establishing data pipelines and analyzers; R.A.D., P.A.B., and S.M. contributed through scoping scientific and strategic direction; H.Q. contributed to technical writing and figure design; K.J.J., F.H.Z., X.W., G.M., H.Z., N.P., J.M.G., and R.G.B. wrote the paper with additional language editing assistance from Claude Opus 4.6 (Anthropic).

**Competing Interest Statement:** The authors are employees of Lila Sciences, Inc. Lila Sciences has filed or may file patent applications related to the materials, methods, or platform described in this work.

# 1 Abstract

Catalyzing acidic oxygen evolution at the proton-exchange-membrane water electrolysis (PEMWE) anode relies almost entirely on iridium or ruthenium, drawn from concentrated supply chains that constrain gigawatt-scale deployment. We report an artificial intelligence (AI)-guided, human-supervised closed-loop platform (>90% automation) integrating combinatorial sputter synthesis, high-throughput screening, machine-learning composition-property models, adaptive multi-objective optimization, and context-aware large-language-model reasoning, where lead catalysts advanced to long-term validation in 1 M $H_2SO_4$ at 10 mA $cm^{-2}$. Navigating a combinatorial metal oxide space, the platform iteratively evaluated the activity-stability trade-off of 2,942 catalysts across 53 material systems and 26 elements, surfacing Ir- and Ru-free complex oxides such as $InMnPdO_x$ and $NiTaPdO_x$ that conventional design logic, and off-the-shelf language models, would not predict. In retrospective benchmarking, our sequential learning agent advanced the activity-stability frontier faster than fixed-policy Bayesian optimization or in-context language-model selection. During long-term testing, $NiTaPdO_x$ operated at lower overpotential than $PdO_x$, but both eventually exceeded 0.5 V: $PdO_x$ at ~200 h and $NiTaPdO_x$ at ~470 h. $InMnPdO_x$ showed a similar overpotential improvement in addition to a dramatic increase in operational stability, retaining overpotential below 0.5 V over 1,000 h of operation. The additive elements promote the formation of a nanostructure that is associated with catalytic activity while stabilizing Pd against corrosion. The results highlight the power of AI-driven science in addressing long-standing challenges in materials chemistry, and the greater availability of Pd relative to incumbent Ir and Ru offers a near-term option to ease supply constraints on scaled electrochemical $H_2$ generation.

# 2 Introduction

The oxygen evolution reaction (OER) in acidic media is the kinetic bottleneck of proton-exchange-membrane water electrolysis (PEMWE) and an enabling half-reaction for emerging electrochemical routes to fuels and chemicals, including $CO_2$ electrolysis and seawater-coupled processes (1, 2). Under the oxidizing, low-pH conditions at the anode, practical devices rely almost exclusively on platinum-group-metal (PGM) oxides of iridium or ruthenium, the only materials that simultaneously deliver adequate activity and durability (3, 4). Despite their scarcity, catalyst cost is not the dominant constraint. System-level analyses show that hydrogen cost is often more sensitive to efficiency and lifetime than to precious-metal loading (5, 6). Resource security, however, remains a deep concern. Iridium ranks among the rarest industrially relevant elements, and ruthenium is only marginally more abundant. Both are recovered as byproducts of PGM mining through a concentrated, inelastic supply chain vulnerable to disruption (7–9). Projected gigawatt-scale electrolyzer deployment amplifies these vulnerabilities, raising the prospect that PGM availability could itself constrain the hydrogen economy (10). Thus, the field needs anode catalysts that draw on elastic supply chains, match Ir- and Ru-class activity, and sustain that activity across the thousands of hours that industrial operation requires; an extremely challenging bar to beat that only a few reported candidates have approached but not met (11, 12).

These intrinsic material challenges are compounded by how catalyst discovery is conducted. Conventional workflows are serial and expert-driven: candidates are chosen from computational priors, literature precedent, or intuition, then synthesized individually and screened in sequence. This approach is slow, labor-intensive, and prone to recycling established material families. High-throughput (HT) experimentation accelerates individual steps, enabling rapid screening of combinatorial libraries for activity and, less often, for stability (13–16). HT activity screening typically runs on the order of seconds per material, leaving extrapolation to long-term durability unresolved. Exhaustive brute-force screening is costly, forcing campaigns to sample the multidimensional composition space conservatively and often without tight feedback between data generation and experiment selection. Both constraints challenge open-ended discovery and leave non-obvious material families to chance. This limitation is especially acute for acidic OER, where activity and long-term durability must be co-optimized. The search space is too large and the durability tests too time consuming to screen exhaustively, so success hinges on selecting the right experiments rather than running more of them (17). Artificial intelligence (AI) and machine learning (ML) are well-suited to this problem by learning from each result to prioritize the next experiment. However, their use in electrocatalysis has stayed predominantly retrospective, interpreting completed datasets rather than prospectively choosing what to run next (18, 19). Closed-loop autonomous laboratories, where a model selects experiments, ingests results, and updates its recommendations, have gained traction across materials science (20–23). Yet, adoption in electrocatalysis remains limited (15), owing to the difficulty of integrating synthesis, characterization, and electrochemical evaluation into one iterative framework that relates diverse descriptors to coupled activity–stability outcomes under harsh conditions (24, 25).

Experimental discovery loops of this kind sit within a broader body of work in AI for science. Deep-learning models trained on large simulated datasets have dramatically expanded the space of candidate materials, as in the discovery of hundreds of thousands of stable inorganic crystals (26). In silico screening, however, applies far less readily to acidic OER. Most electrocatalysts behave as pre-catalysts whose active state emerges only in situ, so activity and dissolution are governed by the reconstructed, charged catalyst–electrolyte interface under operation rather than by the ground-state bulk structures that simulations typically access (27, 28). The field still lacks high-fidelity models of the pre-catalyst-to-catalyst transformation, so AI-guided experimentation that learns the structure–performance map directly from experimental data becomes a practical alternative. Language model-driven agents have meanwhile begun to search for programs and mathematical constructs that surpass established human-designed baselines (29, 30). Closer to the present problem, recent live electrocatalysis loops have used multimodal models mainly for perception and knowledge-guided search-space reduction, optimizing primarily for activity (25).

A fully integrated loop still leaves the central question unresolved: how to choose the next experiment. Large language models (LLMs) are increasingly assigned that choice, though no consensus has emerged on how best to deploy them. At one extreme, LLMs replace the machine-learning surrogates, mapping candidates directly to material properties (31). Alternatively, an LLM serves as a candidate generator, proposing new experiments based on past evaluations (32). Both roles share a limitation. Off-the-shelf LLMs lean on patterns in their training data rather than adapting to experimental feedback (33, 34). So much so that LLM agents working on genetic perturbation and molecular property discovery perform no worse when true outcomes are replaced with randomly permuted labels (35). This failure to explore is most damaging when the target lies in a region of chemical space that literature precedent does not anticipate. Such limitations have motivated hybrid designs that combine LLM reasoning with principled tools or fuse an LLM prior into an active-learning framework (35–38). Whether these strategies transfer to a live, multi-objective experimental loop in acidic OER, where the goal is to discover novel catalysts across a coupled activity–stability space, remains an open question.

In this report, we present an AI-guided, HT, closed-loop platform for the discovery of Ir- and Ru-free metal-oxide electrocatalysts for acidic OER. The platform iteratively maps and learns the global activity–stability landscape of a broad multicomponent oxide space. A sequential learning agent drives this search, shifting between exploration and optimization strategies as the campaign proceeds. The operation is highly automated, with over 90% of lab execution performed robotically and human involvement limited to sample transfers between stations. Three features distinguish this platform from prior autonomous efforts in electrocatalysis. First, it optimizes activity and stability as coupled objectives during HT screening rather than activity alone. Second, it carries the resulting leads beyond screening into long-term validation, confirming durability over more than 1,000 h of operation. Third, by adapting its strategy as evidence accumulated, the sequential learning agent surfaced an entire family of Pd-based oxide catalysts lying outside conventional design precedent, rather than refining performance within an established chemistry.

The two leads, $InMnPdO_x$ and $NiTaPdO_x$, defy expectation on two counts: the relative inactivity of $PdO_x$ (39), steering prior knowledge away from it, and the lack of precedent for Pd-oxide OER catalysts of this kind, in which dilute additives elicit a durable acidic-OER response. $InMnPdO_x$ emerged as the standout of this material family, proving the most durable catalyst in long-term validation against $RuO_x$ and $PdO_x$ benchmarks. Its endurance traces to an operando-formed needle-like nanostructure.

In retrospective benchmarking on our full experimental data, the sequential learning agent advanced the activity–stability frontier faster than any fixed acquisition strategy and reached the Pd family in fewer experiments than a frontier LLM reasoning from prior knowledge. This work shows how prospective, adaptive AI prioritization, integrated with HT experimentation, can expose non-obvious catalyst families and more efficiently navigate that trade-off to identify scalable electrocatalysts.

# 3 Results and Discussion

## 3.1 AI-guided high-throughput platform for closed-loop acidic OER catalyst discovery

To accelerate discovery in the activity–stability-constrained regime of acidic OER, we built a platform that integrates machine-learning prioritization, LLM-assisted decision-making, and modular experimental execution into a unified closed-loop workflow (Fig. 1A and B). The workflow is human-in-the-loop coupled to an AI decision engine: human operators perform sample-fixture transfers and initiate instruments, while experimental runs, data ingestion, model training, ranking, candidate generation, and closed-loop feedback are automated.

In the computational layer (Fig. 1A), uncertainty-calibrated surrogate models are trained on the expanding dataset every round. Composition-based descriptors derived from elemental properties are combined with synthesis metadata, empowering the models to learn simultaneously how composition and processing conditions map to activity (overpotential) and stability (catalyst loss after testing). At each design cycle, an automated model selection procedure evaluates multiple machine learning models and feature representations using cross-validation, selecting the surrogate model with the strongest predictive performance for the coupled activity and stability objectives. The surrogate model is then used by our sequential learning agent, which scores the candidate design space and identifies the highest-priority candidates using acquisition strategies that emphasize exploration, multi-objective optimization, or active search. Finally, an LLM-based reasoning agent selects the most promising experiment from the top-ranked candidates and generates a scientific rationale for the recommendation. Beyond material system and composition, the framework also guides synthesis parameters, annealing temperature, characterization requirements, and progression to long-term durability testing.

Experimentally, the platform links combinatorial sputter synthesis, pre-characterization, rapid acidic OER screening, post-characterization, and selective long-term validation (Fig. 1B). Every combinatorial synthesis batch produces 32 samples, each bearing three distinct oxide composition regions targeted for characterization and testing. These are split across two fixtures of 16 samples (48 unique catalysts per fixture) for the downstream automated workflow. All materials are pre-characterized by optical inspection (OI) for initial quality control and X-ray fluorescence (XRF) for composition and loading. Energy-dispersive X-ray spectroscopy (EDS) is invoked at the direction of the AI pipeline to supplement composition extraction when XRF peak resolution is insufficient. OER screening is performed in 1 M $H_2SO_4$ in a custom-built scanning electrochemical cell (SEC). Activity is taken as the overpotential at 10 mA $cm^{-2}$ after 10 min. Stability is assessed from the overpotential change after an accelerated stress test (AST) together with elemental loss measured before and after screening. Selected catalysts advance to medium-term H-cell testing (≤4 h), and lead candidates to long-term H-cell validation (4–1,000 h), at the same current density. Instrument data was uploaded automatically to a centralized database, where synthesis metadata, composition, loading, electrochemical traces, and derived metrics are standardized for retraining. Screening one 48-catalyst fixture requires less than 10% human hands-on time, with more than 90% of the workflow automated (Fig. 1C): a minimal cycle from synthesis to post-characterization takes around 22 h with an average of around 240 catalysts screened end-to-end per week. ML suggestions were typically generated within minutes.

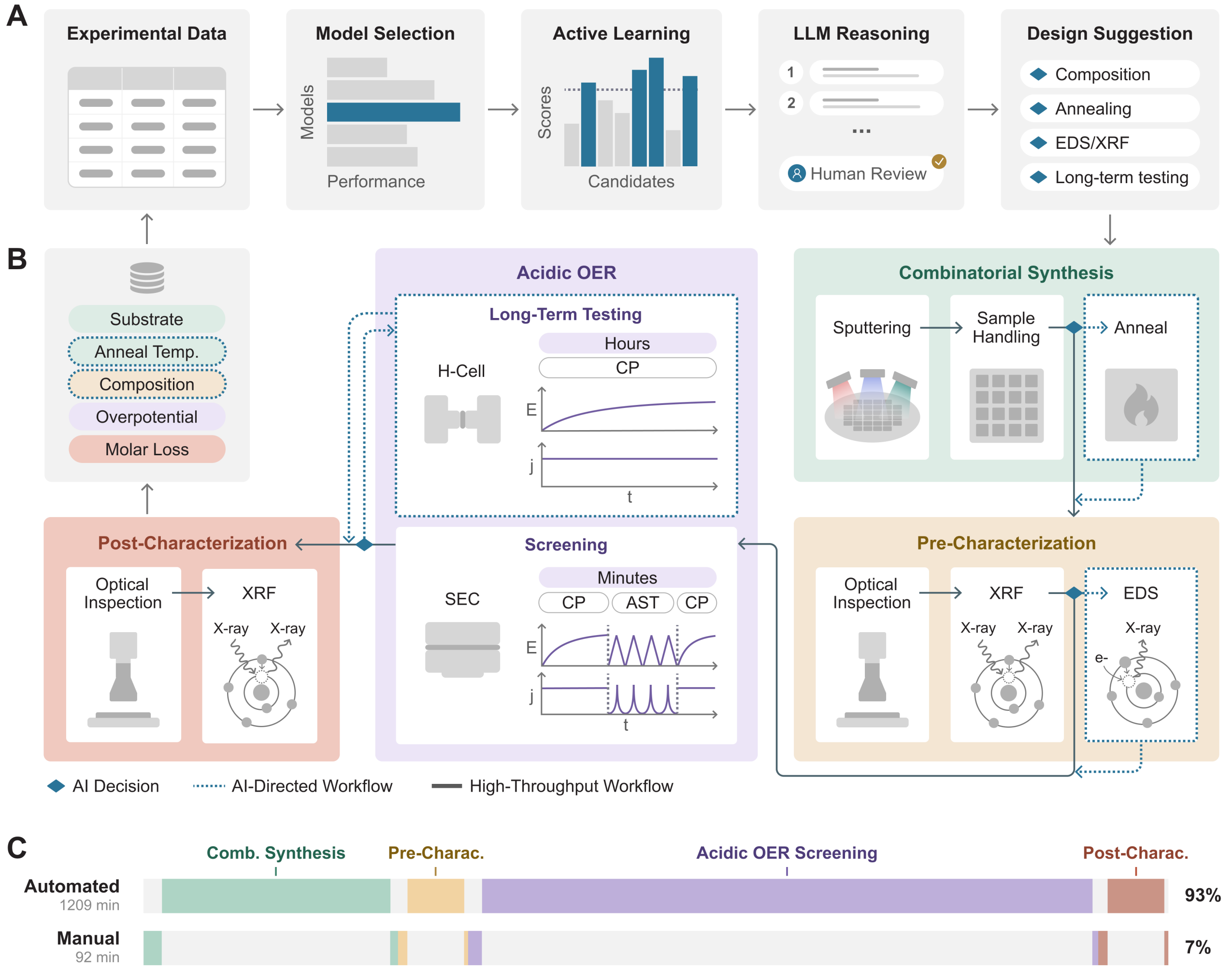

**Fig. 1. Illustration of artificial intelligence (AI)-guided high-throughput (HT) screening platform for acidic oxygen evolution reaction (OER) catalyst discovery.** (A) Schematic of the sequential learning agent, which integrates HT experimentation data, machine learning, and large language model (LLM)-assisted reasoning for iterative catalyst discovery. (B) Experimental pipeline combining combinatorial synthesis, pre-characterization, acidic OER testing, and post-characterization. (C) Time distribution breakdown for a minimum synthesis-to-post-characterization HT workflow of a 48-catalyst batch. Each step comprises manual hands-on and automated portions.

## 3.2 Global activity–stability mapping

To date, the platform has screened 2,942 oxide catalysts spanning 53 material systems and 26 elements (Fig. 2A and B). Down-selection was applied on the fly by excluding catalysts that were analytically unmeasurable or risked damaging the instruments. Projected into activity–stability coordinates, most catalysts fall short on either activity (overpotential at 10 mA $cm^{-2}$) or stability (total molar loss from the electrochemical tests). We therefore summarize performance by the Pareto front, the best attainable trade-offs between overpotential and total molar loading loss. Fig. 2C shows the 1,125 catalysts with overpotential below 0.9 V and measurable XRF molar loss based on sufficient signal-to-noise ratio.

Across the search space, Pd-containing compositions consistently populated the Pareto-optimal regions, despite Pd appearing in only 259 of the 1,125 oxides remaining after filtering. Two material systems stand out in the Pareto front, which are defined by Pd-rich, non-annealed compositions with either In-Mn or Ni-Ta as additives, reaching overpotentials below 0.46 V at 10 mA $cm^{-2}$ with 0.3–0.7 μmol $cm^{-2}$ loss (~60–75% molar retention after screening). Metallic Pd has been reported as an active acidic-OER catalyst through facet strain engineering (40), but Pd-rich catalysts stabilized by dilute additives have not been described as an acidic-OER family, making their emergence here unexpected. We designate the two Pd-rich front-runners, centered near the Ni-Ta and In-Mn Pareto-optimal regions, as $NiTaPdO_x$ and $InMnPdO_x$, respectively.

Several non-PGM systems also exhibited attractive activity within a lower-cost compositional space. The most active were consistently Co-rich and Zr-containing, clustering near $Co_{78}In_{12}Zr_{10}O_x$ and $Co_{86}Fe_7Zr_7O_x$ ($CoInZrO_x$ and $CoFeZrO_x$), reaching 0.55 V at 10 mA $cm^{-2}$. This observation is consistent with Co-based oxides being among the few earth-abundant chemistries shown to catalyze OER in strongly acidic media (41, 42). However, both failed medium-term acidic stability screening, with lifetimes under 4 h (Fig. S1). The Co-Zr non-PGM space thus represents a meaningful improvement but remains bounded by a clear durability ceiling relative to the leading Pd-containing catalysts.

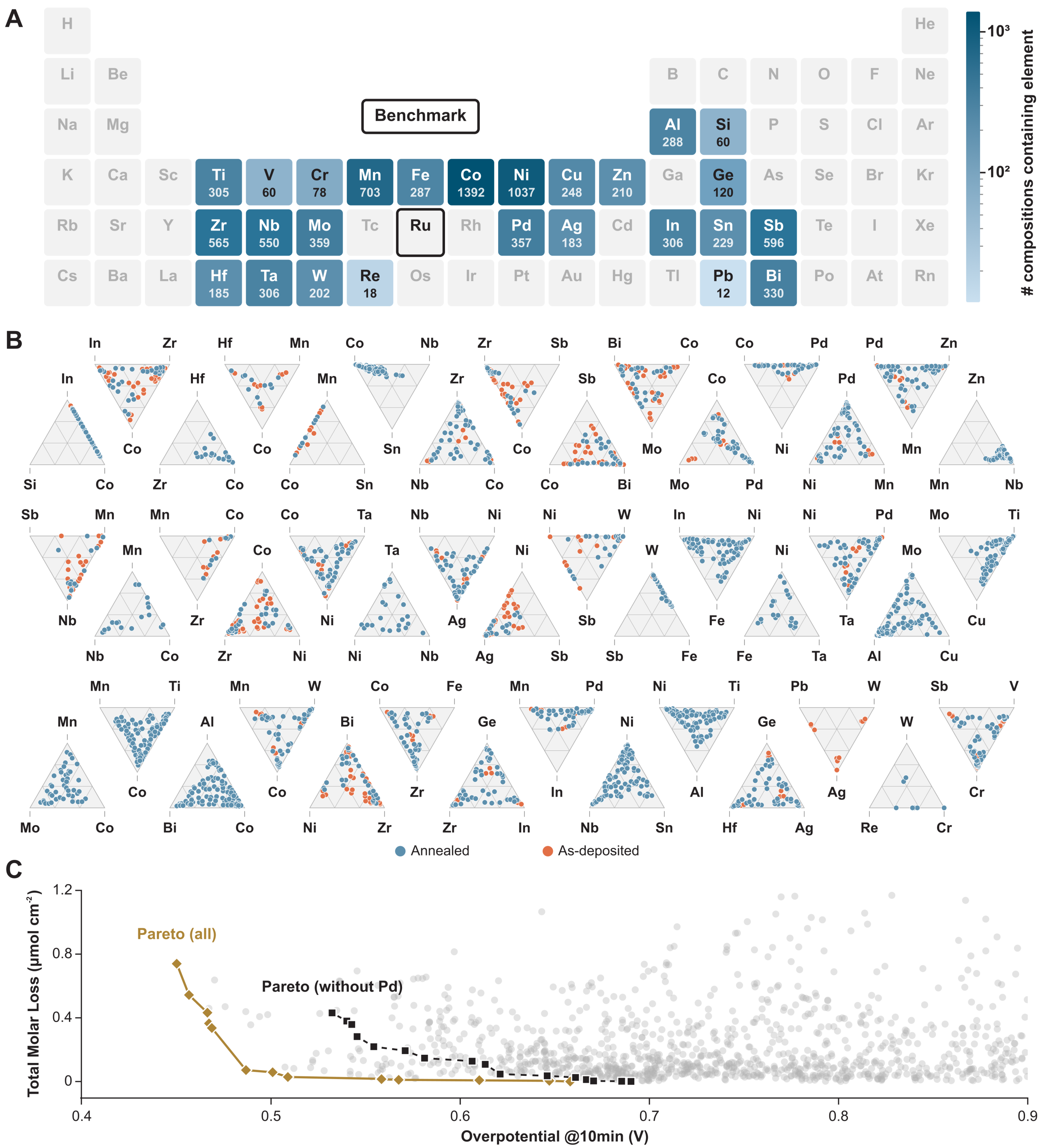

**Fig. 2. HT experimental dataset for acidic OER discovery.** (A) Elemental coverage and frequency contribution of the experimental search space shown on the periodic table. (B) Ternary compositional spaces explored during this study. Colors indicate annealed and as-deposited catalysts. Binary ($BiCoO_x$, $MnSbO_x$) and quaternary ($CoFeHfNiO_x$, $CoFeTaZrO_x$, $CoHfNiSbO_x$, $CoNiSbZrO_x$, $CuNiTaZnO_x$, $MnNbSbSnO_x$, and $MnNbSbTiO_x$) oxide systems are not shown. (C) Multi-objective activity–stability landscape for acidic OER electrocatalysts. Overpotential at 10 mA $cm^{-2}$ after 10 min in 1 M $H_2SO_4$ and X-ray fluorescence (XRF)-derived total molar loading loss were co-minimized to identify Pareto-optimal

catalyst compositions. Gold markers showing Pareto front including Pd-containing catalysts. Black markers showing the Pareto front when excluding Pd-containing catalysts.

### 3.3 Ir- and Ru-free Pd-rich oxides as lead catalysts

Based on the strong performance of the Pd-rich front-runners, we benchmarked the two leads ($InMnPdO_x$ and $NiTaPdO_x$) against $RuO_x$ and $PdO_x$ films made on the same platform and against the strongest non-PGM comparator ($CoFeZrO_x$) found during the screening (Fig. 3). During the initial 10 mA $cm^{-2}$ hold, $InMnPdO_x$ held a stable overpotential of 0.45 V (not iR-corrected), within 40 mV of $RuO_x$ (0.41 V) and roughly 90 mV below the best non-PGM catalyst. $NiTaPdO_x$ and $PdO_x$ behaved similarly (Fig. 3A).

ASTs differentiated the families further. After four cyclic voltammetry (CV) cycles, both Pd-rich leads showed an improvement (decrease) in overpotential in the subsequent 10 mA $cm^{-2}$ chronopotentiometric (CP) measurement, as did the $PdO_x$ benchmark (Fig. 3C), indicating an activation characteristic of Pd-oxide surfaces broadly rather than one induced by the additive elements. Such behavior can arise from reconstruction of under-coordinated Pd sites, roughening by selective dissolution, or formation of a more active surface oxide (43–45). $CoFeZrO_x$, by contrast, degraded under the same sequence.

Pre- and post-screening XRF supported this picture (Fig. 3D). The Pd-containing compositions retained Pd and their minority additives far more completely, whereas $CoFeZrO_x$ lost substantial Co and Zr, consistent with stronger dissolution and the rapid degradation seen during HT and medium-term stability tests. Due to the difficulties in quantifying dilute In and Mn in the Pd-rich matrix during XRF (see SI section 1.2), In and Mn concentrations in Fig. 3D approximate the upper limit of their respective loading. The electrochemical activation and low corrosion of the Pd leads motivated their further study in longer-term testing.

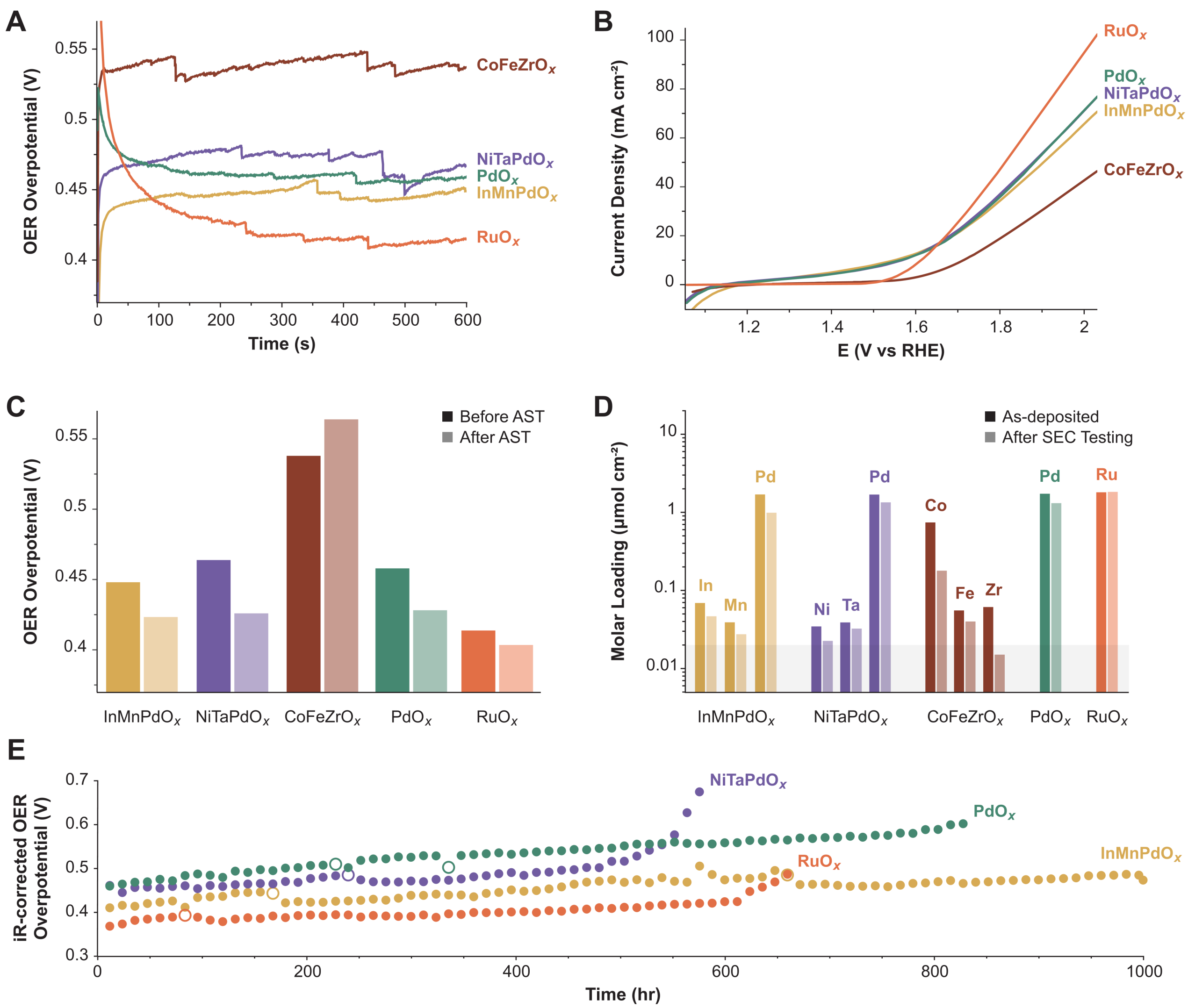

**Fig. 3. Front-runner benchmarking against PdO$_x$ and RuO$_x$.** HT OER screening and long-term testing conducted in 1 M $H_2SO_4$. (A) OER overpotential measured during initial 10 min hold at 10 mA cm$^{-2}$. (B) Anodic sweep of the 2$^{nd}$ cyclic voltammetry (CV) cycle during accelerated stress test (AST). (C) OER overpotential before and after the AST. Bar height represents median of the last 10% of the data in the respective CPs. (D) XRF molar loading determined before and after scanning electrochemical cell (SEC) testing. The In and Mn loadings are calculated from XRF measurements, despite overestimating the relative loadings compared to inductively coupled plasma mass spectrometry (ICP-MS) results (SI section 1.3). Shaded area represents the general 0.02 μmol cm$^{-2}$ detection limit of the XRF. (E) Long-term testing at 10 mA cm$^{-2}$ after iR correction with hollow circles indicating interruption events at which the cell was at open circuit condition. Non iR-corrected data can be found in Fig. S2.

## 3.4 Long-term validation establishes InMnPdO$_x$ as a durable lead catalyst

The two Pd-rich leads, a PdO$_x$ reference, and a RuO$_x$ benchmark were re-synthesized as uniform thin films on the same platinized silicon (Pt/Si) substrate used during HT screening. The target loading for the uniform films was 3.5 - 4 μmol cm$^{-2}$. Screening the uniform films by SEC reproduced the overpotentials of their combinatorial (gradient) counterparts within 15 mV (Fig. S3). Accurately determining the composition of $InMnPdO_x$ is impeded by the low intensity and high attenuation of In and Mn signal in the Pd-rich matrix. Hence, the bulk composition of $InMnPdO_x$ was determined by inductively coupled plasma mass spectrometry (ICP-MS), resulting in 99.91 at.% Pd with a total In-Mn additive of 0.09 ± 0.02 at.% (SI section 1.3). For $NiTaPdO_x$, Ni and Ta are spectrally resolved from Pd, so its composition was quantified reliably by XRF as $Ni_2Ta_2Pd_{96}O_x$. All subsequent long-term testing and characterization were performed on these validated uniform films, in three-electrode H-cells in 1 M $H_2SO_4$ at 10 mA $cm^{-2}$.

Long-term testing was performed as a series of 12-h measurements. Fig. 3E reports the average overpotential for the final hour of each CP, revealing differences in overpotential stability. $PdO_x$ drifts slowly upward across its 828-h window. $RuO_x$ shows a similar early drift followed by a more rapid increase at 615 h, coincident with delamination of the film. $NiTaPdO_x$ holds a steady profile for roughly 500 h before a sharp upward inflection near 520 h. $InMnPdO_x$ is the most stable, sustaining an overpotential below 0.5 V over 1,000 h (repeated measurements over a partial duration on planar and porous supports are shown in Figs. S4 – S6). The stability in overpotential of $InMnPdO_x$ is corroborated by its stabilization of Pd against corrosion. In contrast, PdOx and NiTaPdOx cross the 0.5V mark at ~200 h and ~470 h, respectively. From XRF results, $InMnPdO_x$ retains 96% of its Pd after 1,000 h, a marked improvement in corrosion stability over the other systems. $NiTaPdO_x$ retains 88% Pd after 575 h, $PdO_x$ retains 87% Pd after 828 h, and $RuO_x$ retains 59% Ru after 661 h of operation (Fig. S7).

To understand these stability differences among the Pd-rich films, further in-depth characterization was conducted on films prior to any testing (as-deposited) and after long-term testing (post-test). As-deposited $InMnPdO_x$ and $PdO_x$ showed similar granular top-view morphologies and consistent cross-sectional structures (Fig. 4A and C). X-ray diffraction (XRD) analysis identified polycrystalline tetragonal PdO as the dominant phase in both, with the characteristic (101), (112), and (103) reflections, matching tetragonal PdO reference PDF 00-041-1107 (Fig. 4I). The PdO (101) reflection shifts from 33.90° in $PdO_x$ to 33.92° in $InMnPdO_x$, consistent with lattice contraction upon Mn incorporation, with corresponding shifts observed for the higher angle (112) and (103) reflections. Scherrer analysis of the PdO (101) reflection indicates smaller crystallites in $InMnPdO_x$ (~7.08 nm) than in $PdO_x$ (~10.19 nm) (Table S1). Selected area electron diffraction (SAED) of as-deposited $InMnPdO_x$ shows spotted rings at d-spacings of 2.67, 1.67, and 1.51 Å, matching the (101), (112), (200) planes of tetragonal PdO (Fig. 4G).

After long-term testing, the two films followed different pathways. $PdO_x$ developed aggregated particulates with increased inter-particle spacing while maintaining a consistent cross-section structure (Fig. 4B). Its XRD lost the PdO reflections and gained a new peak near 46.7°, suggesting structural or composition changes during testing (Fig. 4I). $InMnPdO_x$ instead developed a three-layer structure (Fig. 4D, Fig. S8). The surface and intermediate layers both show d-spacings contracted relative to as-deposited $InMnPdO_x$, with domain sizes no larger than 2 nm, appearing as a needle-like surface network over an underlying porous, branched sublayer. The bottom layer mostly retains the grain features (up to 10 nm) and diffraction pattern of the as-deposited material (Fig. 4D, Fig. S9, Table S2). XRD shows reduced intensity of the PdO reflections along with the new peak appearing near 46.7°, consistent with the contracted and distorted lattice of the surface layer and the partially preserved bottom layer seen by SAED (Fig. 4H and I). Post-test X-ray photoelectron spectroscopy (XPS) on the sample surface shows similar chemical states in $InMnPdO_x$ and $PdO_x$ (Fig. S10, Tables S3 and S4), indicating that structural differences rather than surface chemistry govern their divergent durability, consistent with the distinct needle-like nanostructure that forms under operation for $InMnPdO_x$. $NiTaPdO_x$ also developed a dense needle-like nanostructure under electrochemical operation (Fig. 4E and F), suggesting that the additive elements in each system play a role in catalyst restructuring. Differences in stability over hundreds of hours of operation appear to arise from subtle differences in nanostructure.

The needle-like network central to the durability of $InMnPdO_x$ forms in situ, but tailored 1D Pd morphologies can also be grown deliberately. Electrodeposition, for example, converts Pd nanoparticles into nanorods with roughly tenfold higher specific oxygen reduction reaction activity (46). Thus, controlled synthesis of Pd nanostructures followed by oxidation is a plausible route to engineering the active morphology at scale.

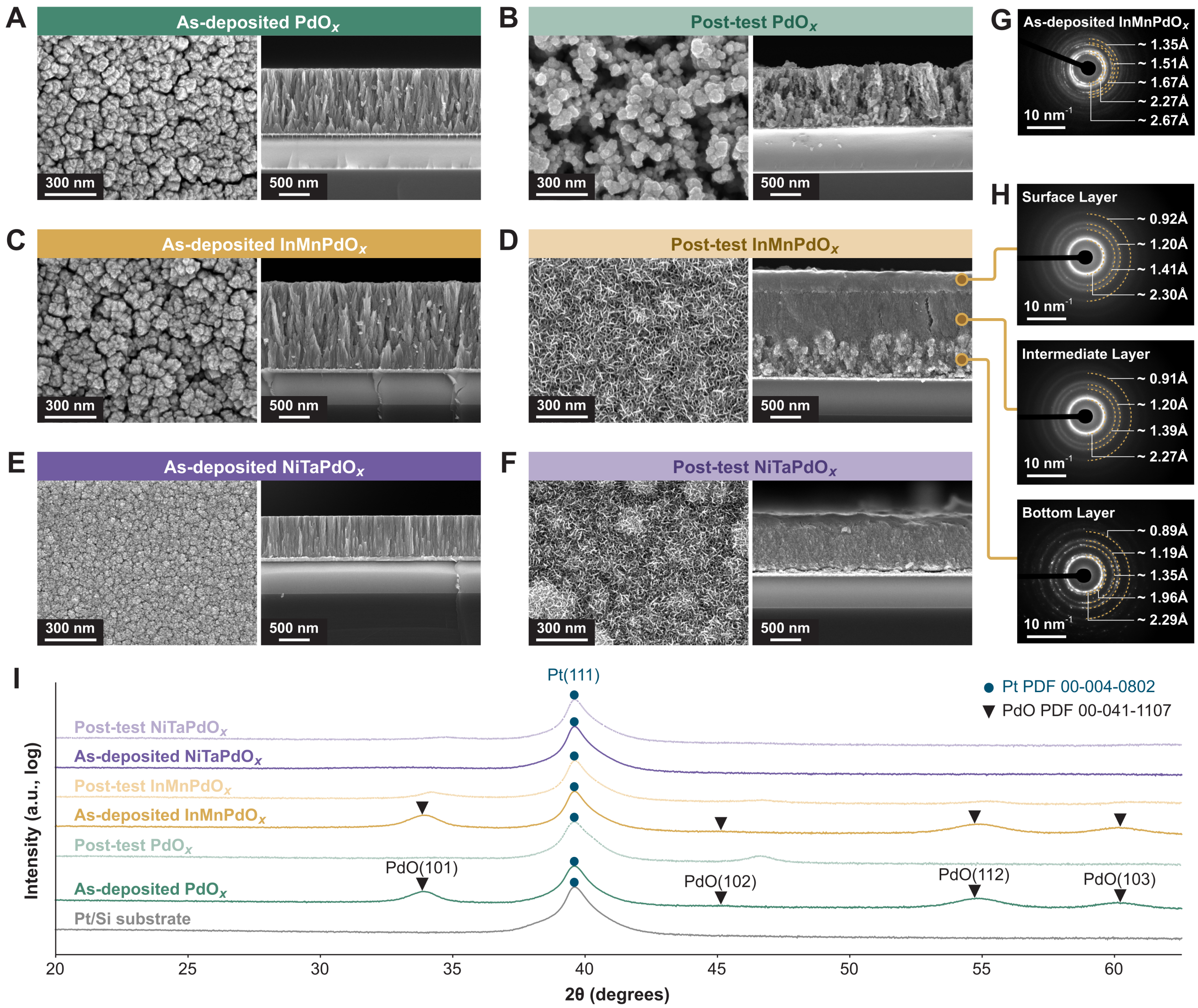

**Fig. 4. Material characterization of Pd-rich front-runners ($InMnPdO_x$ and $NiTaPdO_x$) and benchmark $PdO_x$.** Representative scanning electron microscopy (SEM) top view and cross section view of (A) as-deposited $PdO_x$, (B) post-test $PdO_x$, (C) as-deposited $InMnPdO_x$, (D) post-test $InMnPdO_x$, (E) as-deposited $NiTaPdO_x$, and (F) post-test $NiTaPdO_x$. Representative selected area electron diffraction (SAED) patterns of (G) as-deposited $InMnPdO_x$, and (H) three different regions of post-test $InMnPdO_x$. (I) X-ray diffraction (XRD) spectrum of as-deposited and post-test Pd-rich front-runners and $PdO_x$ benchmark. Complementary XRD of the $RuO_x$ benchmark and of films on a silicon substrate are shown in Fig. S11.

## 3.5 Adaptive strategy selection accelerates OER catalyst discovery

Our AI approach integrates automated machine learning (AutoML) for model and feature selection, uncertainty-aware models for property prediction, and LLM reasoning that adapts the search strategy as evidence accumulates. The agent can choose to focus on multi-objective optimization, which advances the known frontier, or on active search, which targets rare, high-performing regions the frontier does not yet contain. Comparing these methods prospectively would require running a separate live campaign for each, so we benchmarked our approach retrospectively, with all methods selecting from the completed experimental dataset (Fig. 2). The baselines were random search, fixed-policy Bayesian optimization (BO), active search, and an in-context LLM-only strategy. Each baseline isolates one component of our approach: fixed-policy BO and active search commit to a single acquisition strategy, while LLM-only selection removes the surrogate entirely. We selected Pareto hypervolume expansion (trading off overpotential against XRF-derived total molar loading loss) as the target metric. Fixed-policy BO selected experiments by standard expected hypervolume improvement (EHVI); the LLM-only baseline was iteratively prompted with observed results and asked to pick from the remaining candidate pool without surrogate-model uncertainty or formal acquisition. Across multiple seeds, the agent consistently advanced the frontier faster than all baselines, whereas fixed-policy BO converged similarly to random search and LLM-only selection performed even worse than random search (Fig. 5A). The fixed-policy BO result is consistent with the difficulty that mixed categorical–continuous search spaces pose for standard Gaussian-process BO (47, 48), and the weak off-the-shelf LLM performance is consistent with recent reports (33, 37).

In the live campaign, the Pd-oxide family was first reached under the framework's active-search acquisition mode. To determine whether that policy, rather than other policies like multi-objective optimization, was responsible for surfacing this non-obvious material family, we performed a second controlled retrospective comparison. A predictive model trained on the full experimental dataset served as a common oracle: every strategy queried the same model, so differences in performance reflect the selection policy alone. For the LLM baseline we evaluated two variants: 1) with feedback, in which the oracle's prediction for each selected system is returned so the model can adapt subsequent proposals, and 2) without feedback, in which the model proposes from prior knowledge alone. Because such a model cannot be trusted to rank compositions far from its training distribution, we deliberately chose a coarse metric: how many material systems does each strategy need before it selects a Pd-containing catalyst with 90% probability? Reaching the family says nothing about whether a given Pd composition is a good catalyst; that question was addressed with the hypervolume benchmark above, which used only measured results. What this benchmark measures is exploratory reach: whether a strategy arrives at a material system it had no reason to anticipate. The agent reached this threshold in 11 systems and active search in 10, compared with 18 for both fixed-policy BO (Gaussian process (GP) with EHVI) and random search. Both LLM variants (Claude Opus 4.6) failed within the 50-system budget, peaking at a 25% discovery probability with feedback and never selecting the family without it (Fig. 5B). Cumulative discovery probabilities for all strategies are shown in Fig. S12 (SI section 2.1). These results indicate that identification of the Pd family was a reproducible consequence of exploration-driven selection rather than specific experimental ordering.

Across both metrics the agent matched or exceeded the strongest fixed policy: it advanced the activity–stability frontier faster than any of them, and was as reliable as active search (11 vs 10 systems), which was the best fixed policy for discovering the Pd material systems. The strategy selection made by the agent therefore combined the advantage of the best fixed policy on each metric without requiring that policy to be identified in advance.

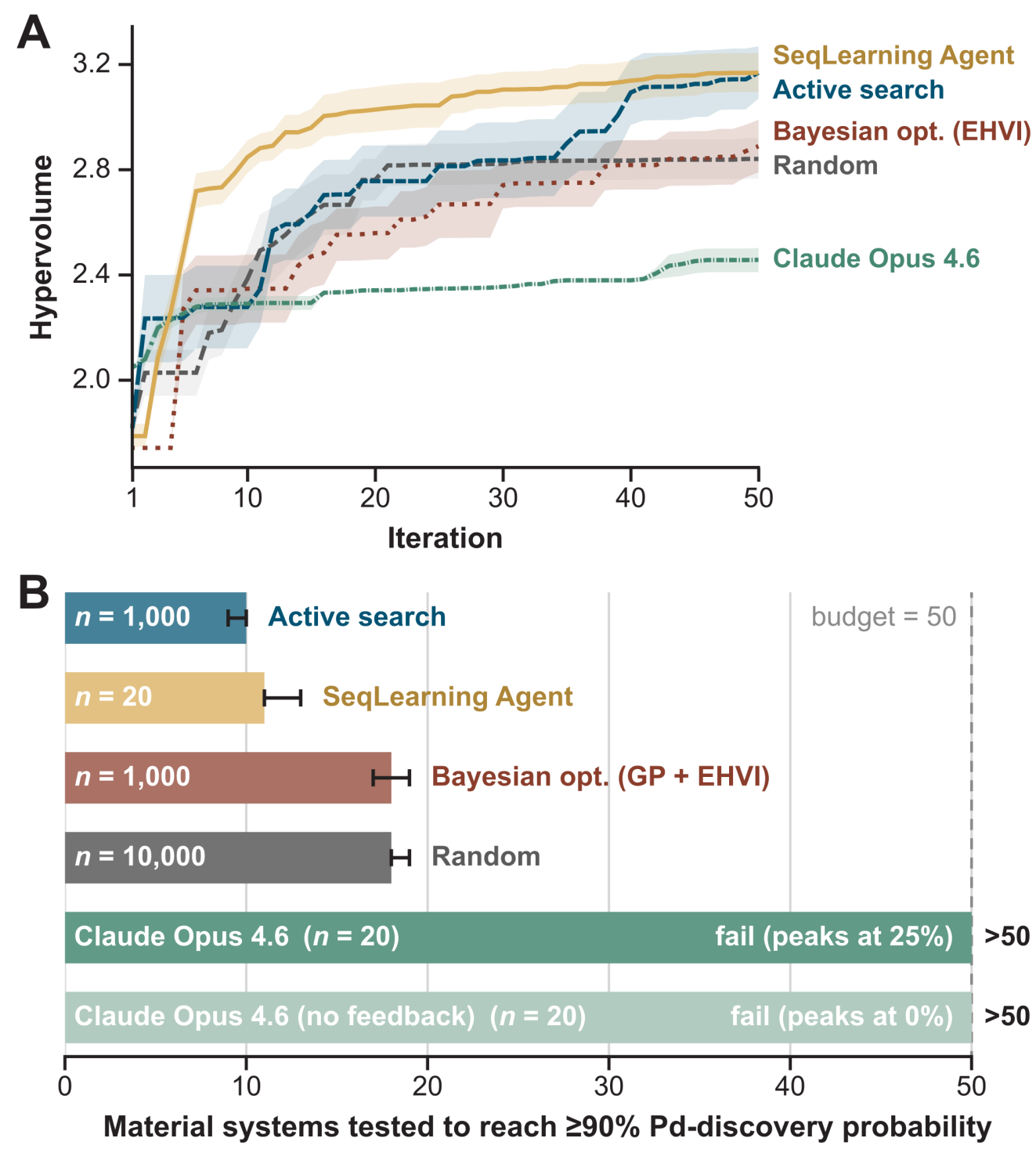

**Fig. 5. Retrospective AI benchmarks of (A) activity–stability optimization, and (B) novelty exploration.** (A) Strategies select from the experimental dataset (Fig. 2) and progress is tracked as activity–stability Pareto hypervolume. Curves are averaged over 10 independent repeats; bands are ±1 standard error of the mean. (B) Strategies select across the full candidate space against a surrogate model trained on that dataset. Bars indicate the material systems needed to select a Pd-containing catalyst in ≥ 90% of repeats; error bars are credible intervals ($5^{th}$–$95^{th}$ percentile) on that crossing point. Strategies spanning the budget mean they never reached the threshold, with the maximum probability attained shown in the bar. EHVI, expected hypervolume improvement; n, independent repeats per strategy; cheaper-to-evaluate strategies are repeated more times.

# 4 Outlook

Although palladium is a PGM, its annual primary supply exceeds iridium's by more than an order of magnitude, its dominant historical demand sink (automotive three-way catalysis) is in structural decline with vehicle electrification, and it is recovered through more diversified mining and recycling channels (49). Pd-rich catalysts therefore relax rather than eliminate the supply constraint at the PEMWE anode, providing a near-term, manufacturable second-source option for gigawatt-scale buildout while the same platform continues to search non-PGM space. The work advances both the catalyst landscape and the methodology by which such families are discovered.

Follow-up studies should focus on resolving the mechanistic origin of the enhanced durability through deeper microscopy and structure–property analysis, with structurally matched controls (notably as-deposited amorphous $PdO_x$ and crystalline as-deposited Ni- or Ta-containing PdO films) that isolate the additive variable from starting crystallinity, alongside rigorous atomic-level simulation. More broadly, the platform could be extended to other electrochemical discovery problems and to richer AI decision-making that includes not only activity and stability, but also cost, supply-chain resilience, and ecological viability, including recycling-aware prioritization.

# 5 Methods

## 5.1 Synthesis

Combinatorial metal oxide thin films were synthesized by reactive magnetron co-sputtering using a Korvus Hex-L physical vapor deposition (PVD) system. Catalyst compositions were identified using the machine learning step of the workflow, drawing compositions from a 26-element search space (Ag, Al, Bi, Co, Cr, Cu, Fe, Ge, Hf, In, Mn, Mo, Nb, Ni, Pb, Pd, Re, Sb, Si, Sn, Ta, Ti, V, W, Zn, Zr). Top-ranked candidates were advanced to the synthesis step for experimental validation. Annealing temperature was selected by AI with a fixed ramp rate of 300 °C per hour and a total anneal time of 5 h. See SI section 1.1 for details.

## 5.2 Characterization

Catalysts were characterized before and after rapid screening. For lead candidates, extensive characterization was conducted on catalysts before and after long-term validation. Composition and loading were measured by XRF, with EDS dynamically added via AI direction where peak overlap limited XRF. Pd-rich front-runners additionally underwent structural and morphological analysis by XRD, scanning electron microscopy (SEM), XPS, transmission electron microscopy (TEM) with SAED, and scanning transmission electron microscopy (STEM). Full instrument and methodology details are in SI Sections 1.2 and 1.3.

## 5.3 Electrochemical screening

HT OER screening in an SEC and H-cell stability testing were both carried out in 1 M $H_2SO_4$ at room temperature. See SI section 1.4 for details on SEC. See SI sections 1.5 and 1.6 for details on H-cell.

## 5.4 Data and experimental records

All measurements, synthesis parameters, and characterization outputs were stored in a centralized database with full provenance. Composition-derived descriptors were linked to experimentally derived activity and stability metrics so that each new result integrated continuously into the learning workflow, keeping outcomes, model predictions, and acquisition decisions traceable across cycles. See SI section 1.7 for details.

### 5.5 Surrogate modeling

Surrogate models for OER activity and stability were trained on composition-derived descriptors and processing variables, with predictive uncertainty from bootstrap ensembles. Because predictive performance evolved as the explored space grew, automated model selection periodically identified the best-generalizing surrogate across multiple architectures. See SI section 1.8 for details.

### 5.6 Candidate selection

Experiment selection was driven by a closed-loop sequential learning agent coupling surrogate models with a portfolio of adaptive active-learning strategies. Candidates were prioritized using an ensemble of acquisition strategies spanning exploitation and exploration, while an LLM-based agent selected the acquisition strategy for each round and tuned its hyperparameters based on optimization progress and accumulated experimental knowledge, targeting improved activity and stability together. See SI section 1.9 for details.

### 5.7 LLM reasoning and benchmarking

Rather than performance predictors, LLMs operated as strategic reasoning agents, selecting the acquisition strategy for each round and thereby modulating the exploration–exploitation balance, prioritizing among acquisition-proposed candidates, and interpreting revealed outcomes from an accumulating, experimentally grounded knowledge base, with human review integrated particularly at early-stage candidate approval. The retrospective benchmark comparing this framework against random search, fixed-policy BO, fixed-policy active search, and LLM-only selection is described in SI Sections 1.10 - 1.12. Claude Opus 4.6 (Anthropic) was used as the benchmark frontier model at the time of this study.

## 6 AI Disclosure

During the preparation of this manuscript, the authors used Claude Opus 4.6 (Anthropic) to improve the writing. The same large language model was used to assist with writing Python scripts for data analysis. All AI-suggested text and code were reviewed by the authors. The authors take full responsibility for the published content.

## 7 Data Availability

Code and data to reproduce the main figures are available at https://github.com/fl97inc/lila_oer_manuscript.

# 8 Acknowledgments

This research was funded entirely by Lila Sciences, Inc. All authors are employees of Lila Sciences and may hold equity or stock options. Lila Sciences has filed or may file patent applications related to the materials, methods, or platform described in this work. The authors would like to thank Francesca Toma for scientific discussions throughout this research. The authors would also like to thank Covalent Corp. for their support in conducting ICP-MS measurements. This work was carried out in part through the use of facilities at MIT.nano.

# 9 References

1. Á. Vass, A. Kormányos, Z. Kószó, B. Endrődi, C. Janáky, Anode Catalysts in CO2 Electrolysis: Challenges and Untapped Opportunities. *ACS Catal.* **12**, 1037–1051 (2022).

2. L. Yu, M. Ning, Y. Wang, C. Yuan, Z. Ren, Direct seawater electrolysis for hydrogen production. *Nat Rev Mater* **10**, 857–873 (2025).

3. C. R. Wang, *et al.*, Proton Exchange Membrane (PEM) Water Electrolysis: Cell-Level Considerations for Gigawatt-Scale Deployment. *Chem. Rev.* **125**, 1257–1302 (2025).

4. C. Wei, *et al.*, Benchmarking Electrocatalyst Stability for Acidic Oxygen Evolution Reaction: The Crucial Role of Dissolved Ion Concentration. *ACS Catal.* **13**, 14058–14069 (2023).

5. R. J. Ouimet, *et al.*, The Role of Electrocatalysts in the Development of Gigawatt-Scale PEM Electrolyzers. *ACS Catal.* **12**, 6159–6171 (2022).

6. M. Hubert, A. M. Esposito, D. Peterson, E. Miller, J. Stanford, “Hydrogen Shot: Water Electrolysis Technology Assessment” (US Department of Energy, 2024).

7. A. Badgett, *et al.*, “Water Electrolyzers and Fuel Cells Supply Chain: Supply Chain Deep Dive Assessment” (US Department of Energy, 2022).

8. N. T. Nassar, T. E. Graedel, E. M. Harper, By-product metals are technologically essential but have problematic supply. *Sci. Adv.* **1**, e1400180 (2015).

9. M. Clapp, C. M. Zalitis, M. Ryan, Perspectives on current and future iridium demand and iridium oxide catalysts for PEM water electrolysis. *Catalysis Today* **420**, 114140 (2023).

10. C. Minke, M. Suermann, B. Bensmann, R. Hanke-Rauschenbach, Is iridium demand a potential bottleneck in the realization of large-scale PEM water electrolysis? *International Journal of Hydrogen Energy* **46**, 23581–23590 (2021).

11. R. Wan, *et al.*, Earth-abundant electrocatalysts for acidic oxygen evolution. *Nat Catal* **7**, 1288–1304 (2024).

12. S. Cherevko, Stabilization of non-noble metal electrocatalysts for acidic oxygen evolution reaction. *Current Opinion in Electrochemistry* **38**, 101213 (2023).

13. J. Huang, *et al.*, Accelerating the Pace of Oxygen Evolution Reaction Catalyst Discovery through Megalibraries. *J. Am. Chem. Soc.* **147**, 30956–30966 (2025).

14. J. M. Gregoire, L. Zhou, J. A. Haber, Combinatorial synthesis for AI-driven materials discovery. *Nat. Synth* **2**, 493–504 (2023).

15. J. M. Przybysz, K. Jenewein, S. Cherevko, High-Throughput Electrochemistry Using Scanning Electrochemical Cells. *ACS Electrochem.* **2**, 801–824 (2026).

16. K. J. Jenewein, G. D. Akkoc, A. Kormányos, S. Cherevko, Automated high-throughput activity and stability screening of electrocatalysts. *Chem Catalysis* **2**, 2778–2794 (2022).

17. A. Ludwig, Discovery of new materials using combinatorial synthesis and high-throughput characterization of thin-film materials libraries combined with computational methods. *npj Comput Mater* **5**, 70 (2019).

18. J. Benavides-Hernández, F. Dumeignil, From Characterization to Discovery: Artificial Intelligence, Machine Learning and High-Throughput Experiments for Heterogeneous Catalyst Design. *ACS Catal.* **14**, 11749–11779 (2024).

19. J. R. Kitchin, Machine learning in catalysis. *Nat Catal* **1**, 230–232 (2018).

20. B. Burger, *et al.*, A mobile robotic chemist. *Nature* **583**, 237–241 (2020).

21. A. G. Kusne, *et al.*, On-the-fly closed-loop materials discovery via Bayesian active learning. *Nat Commun* **11**, 5966 (2020).

22. G. Tom, *et al.*, Self-Driving Laboratories for Chemistry and Materials Science. *Chem. Rev.* **124**, 9633–9732 (2024).

23. N. J. Szymanski, *et al.*, An autonomous laboratory for the accelerated synthesis of inorganic materials. *Nature* **624**, 86–91 (2023).

24. Y. Bai, *et al.*, Stable acidic oxygen-evolving catalyst discovery through mixed accelerations. *Nat Catal* **9**, 28–36 (2026).

25. Z. Zhang, *et al.*, A multimodal robotic platform for multi-element electrocatalyst discovery. *Nature* **647**, 390–396 (2025).

26. A. Merchant, *et al.*, Scaling deep learning for materials discovery. *Nature* **624**, 80–85 (2023).

27. D. Hochfilzer, I. Chorkendorff, J. Kibsgaard, Catalyst Stability Considerations for Electrochemical Energy Conversion with Non-Noble Metals: Do We Measure on What We Synthesized? *ACS Energy Lett.* **8**, 1607–1612 (2023).

28. H. Xin, *et al.*, Roadmap for transforming heterogeneous catalysis with artificial intelligence. *Nat Catal* **9**, 102–111 (2026).

29. B. Romera-Paredes, *et al.*, Mathematical discoveries from program search with large language models. *Nature* **625**, 468–475 (2024).

30. A. Novikov, *et al.*, AlphaEvolve: A coding agent for scientific and algorithmic discovery. [Preprint] (2025). Available at: https://arxiv.org/abs/2506.13131 [Accessed 14 September 2026].

31. M. Caldas Ramos, S. S. Michtavy, A. D. White, M. D. Porosoff, Bayesian Optimization of Catalysis with In-Context Learning. *ACS Cent. Sci.* **12**, 599–615 (2026).

32. C. Yang, *et al.*, Large Language Models as Optimizers in *The Twelfth International Conference on Learning Representations*, (2024).

33. A. Kristiadi, *et al.*, A Sober Look at LLMs for Material Discovery: Are They Actually Good for Bayesian Optimization Over Molecules? in *Proceedings of the 41st International Conference on Machine Learning*, Proceedings of Machine Learning Research., R. Salakhutdinov, *et al.*, Eds. (PMLR, 2024), pp. 25603–25622.

34. A. Krishnamurthy, K. Harris, D. J. Foster, C. Zhang, A. Slivkins, Can large language models explore in-context? in *Proceedings of the 38th International Conference on Neural Information Processing Systems*, NIPS ’24., (Curran Associates Inc., 2024).

35. R. Gupta, J. Hartford, B. Liu, LLMs for Bayesian Optimization in Scientific Domains: Are We There Yet? in *Findings of the Association for Computational Linguistics: EMNLP 2025*, C. Christodoulopoulos, T. Chakraborty, C. Rose, V. Peng, Eds. (Association for Computational Linguistics, 2025), pp. 15482–15510.

36. D. Agarwal, *et al.*, Searching for Optimal Solutions with LLMs via Bayesian Optimization in *The Thirteenth International Conference on Learning Representations*, (2025).

37. M. Akke, *et al.*, Bayesian Optimization for Biochemical Discovery with LLMs. [Preprint] (2025). Available at: https://chemrxiv.org/doi/abs/10.26434/chemrxiv-2025-w1wsh [Accessed 16 September 2026].

38. T. Liu, N. Astorga, N. Seedat, M. van der Schaar, Large Language Models to Enhance Bayesian Optimization in *The Twelfth International Conference on Learning Representations*, (2024).

39. S. Cherevko, *et al.*, Dissolution of Noble Metals during Oxygen Evolution in Acidic Media. *ChemCatChem* **6**, 2219–2223 (2014).

40. J. Peng, *et al.*, Hierarchical palladium catalyst for highly active and stable water oxidation in acidic media. *National Science Review* **10**, nwac108 (2023).

41. J. S. Mondschein, *et al.*, Crystalline Cobalt Oxide Films for Sustained Electrocatalytic Oxygen Evolution under Strongly Acidic Conditions. *Chem. Mater.* **29**, 950–957 (2017).

42. L. Chong, *et al.*, La- and Mn-doped cobalt spinel oxygen evolution catalyst for proton exchange membrane electrolysis. *Science* **380**, 609–616 (2023).

43. P. Strasser, Free Electrons to Molecular Bonds and Back: Closing the Energetic Oxygen Reduction (ORR)–Oxygen Evolution (OER) Cycle Using Core–Shell Nanoelectrocatalysts. *Acc. Chem. Res.* **49**, 2658–2668 (2016).

44. J. Liu, L. Guo, In situ self-reconstruction inducing amorphous species: A key to electrocatalysis. *Matter* **4**, 2850–2873 (2021).

45. L. D. Burke, J. K. Casey, The electrocatalytic behaviour of palladium in acid and base. *J Appl Electrochem* **23**, 573–582 (1993).

46. L. Xiao, L. Zhuang, Y. Liu, J. Lu, H. D. Abruña, Activating Pd by Morphology Tailoring for Oxygen Reduction. *J. Am. Chem. Soc.* **131**, 602–608 (2009).

47. B. Ru, A. Alvi, V. Nguyen, M. A. Osborne, S. Roberts, Bayesian Optimisation over Multiple Continuous and Categorical Inputs in *Proceedings of the 37th International Conference on Machine Learning*, Proceedings of Machine Learning Research., H. D. III, A. Singh, Eds. (PMLR, 2020), pp. 8276–8285.

48. X. Wan, *et al.*, Think Global and Act Local: Bayesian Optimisation over High-Dimensional Categorical and Mixed Search Spaces in *Proceedings of the 38th International Conference on Machine Learning*, Proceedings of Machine Learning Research., M. Meila, T. Zhang, Eds. (PMLR, 2021), pp. 10663–10674.

49. A. Cowley, *et al.*, “PGM market report” (Johnson Matthey, 2026).

# Supporting Information

## AI-guided high-throughput discovery of iridium- and ruthenium-free palladium-oxide catalysts for durable acidic oxygen evolution

Ken J. Jenewein, Faezeh Habib Zadeh, Xiaoxiao Wang, Gustavo Malkomes, Huafan Zhang, Natalie Page, Jae Jin Bang, Peter J. Santiago, Karla V. Contreras, Katherine K. Li, Allison Perna, Lorena M. Britton, Fahrettin Kilic, Kevin J. Cruse, Armin Taheri, Krishnanand Mallayya, Harley Quinn, Rebecca A. Durr, Peter A. Beaucage, Santiago Miret, John M. Gregoire, Rafael Gómez-Bombarelli

Email: jgregoire@lila.ai and rgbombarelli@lila.ai

# 1 Supporting Methods

## 1.1 Film synthesis

Combinatorial metal oxide thin films were synthesized by reactive magnetron co-sputtering using a Korvus Hex-L physical vapor deposition (PVD) system. Thin films were deposited on 2 × 2 cm substrates including silicon, platinum-coated silicon with titanium adhesion layer (Pt/Si), and platinum-coated titanium porous transport layer (PTL). Substrates were mounted to a custom 150 mm diameter stainless steel fixture which holds 32 substrates and masked with a custom aluminum mask covering one edge of every substrate up to 3.2 mm. Masking the substrate prior to deposition results in a step edge for electrical contact to the substrate for electrochemical testing. The chamber was evacuated to a base pressure of $<7.5\times10^{-7}$ Torr with a Pfeiffer HiPace 700 turbo-pump backed by an Agilent IDP-7. Substrates were bias-etched using a Seren RF power supply at 100 W for 10 min to remove any surface contamination and improve coating adhesion. The deposition pressure for all synthesis was kept constant at 12.9 mTorr with a gas flow rate of 40 sccm Ar and 10 sccm $O_2$ for an oxygen partial pressure of 20%. Combinatorial synthesis during the high-throughput workflow was performed with a substrate rotation of 0 RPM to form a gradient of metal oxide alloys across the 150 mm substrate fixture. Catalysts for long-term performance validation were deposited as uniform thin films with a substrate rotation of 20 RPM. Thin films were deposited from three or four 50.8 mm diameter metal targets with purity ≥99.9%. All ternary thin films were sputtered using three magnetron cathodes positioned equidistantly and pointed at the fixture edge with a working distance of approximately 85 mm. Quaternary thin films were sputtered using a fourth cathode positioned between two of the equidistant cathodes and pointed toward the center of the fixture with a working distance of approximately 169 mm. The cathodes were controlled by Seren RF power supplies at powers ranging from 3–60 W and were pre-sputtered for 2 min prior to opening the substrate shutter and starting the deposition. Powers were selected using an internal PVD predictor model and theoretical sputter rates to target top ranked candidate compositions in the center of the 150 mm diameter combinatorial spread, to synthesize the targeted composition and the surrounding compositions. The deposition time ranged from 3–5 h for gradient depositions and 20–23 h for uniform coatings. Uniform coatings were deposited to thicknesses equaling a molar loading around 3.5–4 μmol $cm^{-2}$. A subset of the sputtered samples was transferred to custom Macor furnace fixtures and annealed in air in a Thermo Fisher Thermolyne or Lindberg Blue/M furnace at a fixed ramp rate of 300 °C per hour and a total anneal time of 5 h. The annealing temperature was dictated by the sequential learning agent.

## 1.2 Characterization during high-throughput workflow

Characterization on as-deposited films included optical inspection (OI), X-ray fluorescence (XRF), and optional energy-dispersive X-ray spectroscopy (EDS) mapping. OI images were captured using a custom setup, consisting of a Thorlabs Kiralux camera, automated stage, and a Novteke white light source. XRF spectra were measured by a Bowman P-series energy-dispersive XRF (EDXRF) system, employing a Rhodium X-ray source (50 kV, 1 mA) with no primary filter for maximized signal intensity for the broad range of element selection. The XRF system was calibrated by commercial calibration standards from Micromatter Technologies Inc., for elemental mass and molar loading quantification. The compositions used as inputs for the closed-loop learning were extracted from either XRF or EDS, decided by the artificial intelligence (AI) pipeline developed in this study, which selects the appropriate technique capable of producing better-resolved spectral peaks. If EDS is selected during the high-throughput workflow, a Thermo Fisher Scientific Phenom XL G2 scanning electron microscope (SEM) was used to conduct the measurements. XRF, SEM, and EDS of as-deposited samples were taken at the center of each catalyst region (three regions per 20 × 20 mm substrate).

Post-test characterization includes OI and XRF measurements using the same setup with the same instrument parameters, except for the post-XRF measurements on the H-cell tested samples, where a 5 × 5 measurement grid covering the entire tested region was applied to better capture the post-test loading variations on the sample surface.

Especially for the $InMnPdO_x$ front-runner, low In and Mn intensities cause high uncertainty when quantifying the composition in a Pd-rich matrix. Accordingly, the bulk composition of the $InMnPdO_x$ front-runner was determined independently by inductively coupled plasma mass spectrometry (ICP-MS; SI section 1.3). XRF-derived In and Mn values across the broader screening library are therefore treated as comparative rather than absolute.

## 1.3 Characterization during validation step

Because of the difficulties in accurately quantifying In and Mn concentrations in a Pd-rich matrix using XRF or EDS for $InMnPdO_x$ (SI section 1.2), the bulk composition of the uniform $InMnPdO_x$ films was characterized by Covalent Corp. (CA, United States) using a Thermo Fisher Triple-Q ICP-MS. Samples were digested overnight in freshly prepared aqua regia in a pre-cleaned PFA vessel with hot plate heating. After cooling, each extraction was quantitatively transferred into a volumetric flask, brought to volume with ultra-pure water, and further diluted as needed. The diluted samples were analyzed in kinetic energy discrimination (KED) mode. Sc and Y were used as internal standards to monitor matrix-related space-charge effects and correct for instrumental drift. Averaged across three films from the validation batch, the composition was 99.91 ± 0.01 at.% Pd, 0.06 ± 0.01 at.% Mn, and 0.03 ± 0.01 at.% In. Ni and Ta are spectrally resolved from Pd, so XRF composition is reliable and independent confirmation was not required for $NiTaPdO_x$.

Extensive morphology and structural characterization were conducted before long-term H-cell tests. X-ray diffraction (XRD) measurements were performed using the Bruker D8 diffractometer with an incident Cu Kα beam at a fixed incident angle of 20 degrees and measured by a 2D Eiger 500K detector. XRD results were analyzed by the Bruker EVA application. SEM images of the best-performing Pd-containing thin-film samples were acquired using a Thermo Fisher Helios 5 system (5 kV and 0.2 nA). Transmission electron microscopy (TEM) with selected area electron diffraction (SAED) and scanning transmission electron microscopy (STEM) images were obtained from Thermo Fisher Talos F200i system at MIT.nano (MA, United States). As-deposited and post-test $InMnPdO_x$ lift-out lamella were prepared by a Thermo Fisher Helios 5 system. X-ray photoelectron spectroscopy (XPS) measurements were conducted using PHI Versaprobe II at MIT.nano (Al Kα source, 15 kV, 200 μA) with electron neutralization. The data was analyzed by CasaXPS software (1). All spectra were fitted with a combination of Gaussian–Lorentzian (GL) line shapes with a Shirley-type background subtraction. Binding energy positions were referenced to the C1s C—C peak at 284.8 eV. Pd3d peaks from as-deposited films can be fitted by Pd(0) (metallic Pd) peaks, PdO (stoichiometric oxide) peaks, defective PdO peaks, and Pd satellite peaks (introduced by metallic Pd). Pd3d peaks from post-tested samples can be fitted by PdO (stoichiometric oxide) peaks, defective PdO peaks, and $PdO_x$ (high binding energy oxidation state) (2, 3). For Pd(0), an asymmetric Lorentzian line shape LA(1.9,7,2) was used to account for metallic screening effects, whereas oxidized Pd species were fitted using symmetric GL(30) line shape. No fixed intensity constraints were imposed between different chemical states, except for spin–orbit splitting and area ratios within each doublet. All primary spin–orbit split doublets were fixed with an energy separation of 5.26 eV and an area ratio of 3:2 ($3d_{5/2}$ : $3d_{3/2}$). To account for slight asymmetry in line broadening between spin–orbit components arising from inelastic scattering, background subtraction effects, and final-state screening contributions in transition-metal oxides, a constrained full width at half maximum (FWHM) relationship was applied within all oxide-related chemical states, such that FWHM($3d_{3/2}$) was allowed to vary within 1.0–1.2× FWHM($3d_{5/2}$) (4). Shake-up satellites were fitted with the spin-orbit separation fixed at 5.26 eV and a constant 2.70 eV offset from the parent component, with the area ratio left unconstrained. Satellite components are excluded from the normalized distributions in Table S4.

## 1.4 Rapid electrochemical screening

High-throughput (HT) oxygen evolution reaction (OER) activity screening was performed using an scanning electrochemical cell (SEC) adapted from a previously reported design (5). The anode and cathode compartments were separated by a bipolar membrane (Fumasep FBM). A leak-free Ag/AgCl electrode (sat. KCl, Innovative Instruments) served as the reference electrode, and a Pt wire (0.1 mm diameter, 99.99%) was used as the counter electrode. The reference electrode was calibrated versus the reversible hydrogen electrode (RHE, HydroFlex, Gaskatel) at the start of each run. All measurements were carried out in 1 M $H_2SO_4$ (pH 0.3) at room temperature using a Gamry Reference 620 potentiostat. Instrument control and workflow orchestration were implemented through a custom HELAO-async framework (6). The HT electrochemical protocols used in SEC screening comprised three sequential steps. First, each catalyst was subjected to a 10 min chronopotentiometric (CP) hold at 10 mA $cm^{-2}$. This was followed by four cyclic voltammetry cycles between 1.0 and 2.0 V versus RHE at a scan rate of 125 mV $s^{-1}$, which served as an accelerated stress test (AST) intended to simulate non-steady-state operation. The protocol concluded with a second CP hold at 10 mA $cm^{-2}$ for 1 min to assess the effect of the AST on catalyst performance. To increase screening throughput, an early-stopping criterion was implemented where measurements were terminated if the cell reached instrument overload or if the catalyst exhibited an overpotential above the substrate after 180 s of the initial CP hold. Electrochemical surface area (ECSA) determination and iR correction were not incorporated into the high-throughput screening protocol. The workflow was designed as a comparative discovery screen under fixed geometric and electrolyte conditions with throughput and internal consistency across thousands of measurements that took precedence over absolute activity metrics. More detailed electrochemical characterization was reserved for downstream validation of lead candidates.

To monitor instrument health throughout the study, quality-control measurements on fluorine-doped tin oxide (FTO) were performed between successive thin-film samples. Reproducibility of the SEC protocol was assessed from five bare Pt/Si substrates measured at 10 mA $cm^{-2}$ and yielded a variation of ±13 mV with a mean overpotential of 0.93 V.

## 1.5 Long-term durability

All long-term electrochemical measurements were performed on Gamry Interface 1010E potentiostats. Each catalyst was tested in a custom three-electrode glass H-cell with a Nafion 115 membrane separating the working- and counter-electrode compartments. The working electrode was a thin-film catalyst coating on a Pt/Si carrier substrate (3 $cm^2$ geometric area exposed to electrolyte); the counter electrode was a Pt foil (15 × 15 × 0.1 mm, 99.99%); and the reference electrode was a leak-free Ag/AgCl electrode (sat. KCl, Innovative Instruments) positioned in the working-electrode compartment. The electrolyte was 1 M $H_2SO_4$ (pH 0.3, measured at the start of each run). All tests were performed at room temperature. Each reference electrode was calibrated versus RHE at the start of each run.

Each long-term run was constructed by repeating a four-step sequence: a 60 s open circuit potential (OCP) measurement at the start of each cycle, a 43,200 s CP hold at 10 mA $cm^{-2}$, a 30 s OCP measurement at the end of cycle, and a potentiostatic electrochemical impedance spectroscopy (PEIS) measurement at OCP. The PEIS per cycle was acquired with a 10 mV root-mean-square AC perturbation across 0.1 Hz to 10 kHz at 10 points per decade. All PEIS spectra were fit to a series-resistance ($R_s$) plus constant-phase element (CPE) equivalent circuit.

The cumulative CP overpotential trace for each cell was iR-corrected at each operating point by subtracting the time-matched ohmic drop:

$$\eta_{corrected}(t) = E_{vs.RHE}(t) - 1.229\ V - i \cdot R_s(t)$$

with $i$ = 0.030 A (the constant galvanostatic current on every cell) and $R_s(t)$ extracted by fitting the PEIS at each cycle.

## 1.6 Replicate cells

Replicate H-cell runs were prepared with samples from the same sputtering batch as each long-term cell reported in Fig. 3E, mounted under identical conditions, and operated under the same 12 h cycling protocol. Additionally, for $InMnPdO_x$, a nominally duplicate thin-film sample on PTL was surveyed under the same long-term conditions. Replicate CP traces were iR-corrected using each replicate's own PEIS $R_s(t)$.

## 1.7 Experimental records and descriptors

All experimental measurements, synthesis parameters, and characterization outputs were stored in a centralized structured database designed to support iterative closed-loop optimization. Each experimental record maintained full provenance, including catalyst composition, annealing conditions, substrate identity, electrochemical measurements, and post-test analysis. Composition for each catalyst was taken from XRF or, where peak overlap limited XRF quantification, from EDS. Composition-dependent descriptors were computed directly from elemental identities and linked to experimentally derived activity and stability metrics, enabling continuous integration of newly generated data into the learning workflow. This structure allowed experimental outcomes, model predictions, acquisition decisions, and subsequent analyses to remain fully traceable across sequential optimization cycles.

## 1.8 Surrogate models and model selection

The sequential learning agent integrates automated surrogate modeling with adaptive multi-objective acquisition strategies to guide sequential learning of the catalyst design space. Predictive models for acidic OER activity and stability were trained using composition-derived descriptors together with experimentally relevant processing variables. Because predictive performance evolved as the explored compositional space expanded, automated model selection was performed periodically across multiple regression architectures to identify the surrogate model with the strongest cross-validated generalization performance at each stage of the campaign. Predictive uncertainty was estimated using bootstrap ensemble methods, enabling probabilistic evaluation of unexplored candidates.

## 1.9 Adaptive acquisition and multi-objective optimization

The sequential learning agent prioritized candidates using a porfolio of acquisition strategies, including expected hypervolume improvement (EHVI), Thompson sampling over surrogate posterior distributions, stochastic random exploration to preserve diversity, and a lookahead acquisition function. The lookahead strategy was designed to account for the future value of information during sequential optimization. Candidate experiments were first pre-screened using a computationally efficient probability-of-success metric based on the joint likelihood of meeting predefined activity and stability targets. For each shortlisted candidate, the surrogate model was hypothetically updated using the candidate's predicted activity and stability values, and the updated model was subsequently used to evaluate the remaining design space. Candidate scores combined the probability of immediate success with the expected utility of the resulting model update, favoring experiments that were both intrinsically promising and expected to improve future optimization decisions.

By integrating multiple acquisition strategies rather than relying on a single criterion, the framework balanced exploitation of high-performing regions with continued exploration of uncertain or under-sampled areas of compositional space. The multi-objective optimization jointly minimized overpotential and XRF-derived material loss. Beyond composition selection, the framework also informed decisions related to synthesis conditions, annealing protocols, characterization requirements, and progression to extended H-cell testing. Experimental outcomes from each iteration were incorporated back into the training dataset, enabling continuous updating of surrogate models, acquisition priorities, and optimization strategy throughout the campaign.

### 1.10 LLM reasoning layer

Large language models (LLMs) were incorporated as reasoning agents within the sequential learning agent to provide adaptive strategic guidance complementary to quantitative surrogate modeling. The LLMs did not directly predict catalytic performance; instead, they operated at higher levels of experimental decision-making, including adjustment of exploration–exploitation behavior, prioritization among candidate experiments proposed by acquisition functions, and interpretation of experimentally revealed outcomes. Optimization decisions were informed not only by model predictions and uncertainty estimates, but also by accumulated contextual knowledge regarding optimization dynamics, model reliability, and experimentally observed compositional trends. To support iterative reasoning across sequential optimization cycles, the framework maintained persistent knowledge structures populated exclusively with experimentally grounded observations generated after candidate evaluation. These knowledge representations captured both optimization-level heuristics and materials-domain insights, enabling the system to accumulate empirical context over the course of autonomous exploration rather than treating each iteration independently. Together, the surrogate models, adaptive acquisition strategies, and LLM reasoning layer constitute the sequential learning agent that directs the closed-loop discovery campaign.

### 1.11 Retrospective benchmark of selection strategies

The performance of the agentic sequential learning framework was evaluated retrospectively using the experimentally generated dataset as a fixed candidate environment for sequential selection. The benchmark task consisted of iteratively selecting catalyst compositions and annealing conditions to maximize Pareto hypervolume expansion across the joint activity–stability objective space, with both overpotential and XRF-derived material loss minimized. Comparator methods included fixed-policy strategies (active search, random exploration, and a conventional Bayesian optimization with Gaussian process and selecting by expected hypervolume improvement) and an LLM-only iterative selection strategy that chose from observed results and the remaining candidate pool without surrogate-model uncertainty estimation or formal acquisition. Performance was quantified by the rate of Pareto hypervolume expansion as a function of experimental iterations, with mean and standard error derived from 10 repeated runs using different random seeds (Fig. 5A). Although retrospective benchmarking cannot fully capture all dynamics of a live experimental campaign, this framework enabled controlled comparison of optimization strategies under identical candidate pools and objective landscapes. Notably, this replay approach also implicitly favors strategies by including catalysts discovered by the original campaign, an advantage that would not hold for a live, prospective search.

### 1.12 Retrospective benchmark of novelty exploration

To compare selection policies free of the pool bias inherent to the replay benchmark of SI section 1.11, we considered the candidate space of all three-element combinations of 25 of the 26 sputter-available elements. Re is the omitted element: it appears in a single material system in our dataset (Cr-Re-W), leaving the surrogate no basis for ranking Re-containing candidates. Outcomes were scored against a surrogate oracle in place of measurements: an XGBoost model trained on campaign data. The surrogate is informative only near the compositions it was trained on and cannot be trusted to predict performance far from that distribution, so we pose a simpler question: how quickly does each strategy first propose a Pd-containing system? Policies are compared by the number of systems needed to reach ≥ 90% probability of having made such a selection. At each step a strategy proposed a three-element system, and each run terminated at its first Pd-containing selection, within a budget of 50 systems (Fig. S12). Two language-model variants were evaluated as well. The "with feedback" variant received the oracle's prediction for each system it selected, accumulated in its context before the next query. The "no feedback" variant proposed from prior knowledge alone, seeing only its own earlier suggestions so as to avoid repeats, serving as an unaided test of what the model's priors recommend.

# 2 Supporting Discussion

## 2.1 Pd-discovery speed across selection policies

The replay benchmark of Fig. 5A draws candidates from the pool our campaign collected, which is enriched toward the regions the live strategy explored. The direct-querying benchmark of Fig. 5B removes that bias by scoring every strategy against a common surrogate oracle over the full space of three-element systems (SI section 1.12).

The coarseness of the target flatters random search: it explores broadly by construction, and with 276 of the 2,300 possible systems (12.0%) containing Pd, chance alone can find one quickly. Even so, active search, the sequential learning agent, and fixed-policy Bayesian optimization found Pd in every run, whereas neither language-model variant reached the 90% target within the budget. With feedback, a Pd-containing system was selected in 5 of 20 runs (25%, the peak probability in Fig. 5B), in every case only in the last few steps of the budget (steps 46–48); no feedback, in 0 of 20 runs, against 9,987 of 10,000 for random search.

This benchmark is deliberately simplified. It treats all Pd-containing catalysts as equally valuable discoveries, which is not physically realistic, and it omits the selection of specific compositions within each material system, in order to keep the task and its context tractable for the language models. Reported probabilities should therefore be read as a relative comparison of selection policies under controlled conditions rather than as absolute measures of discovery performance.

# 3 Supporting Figures

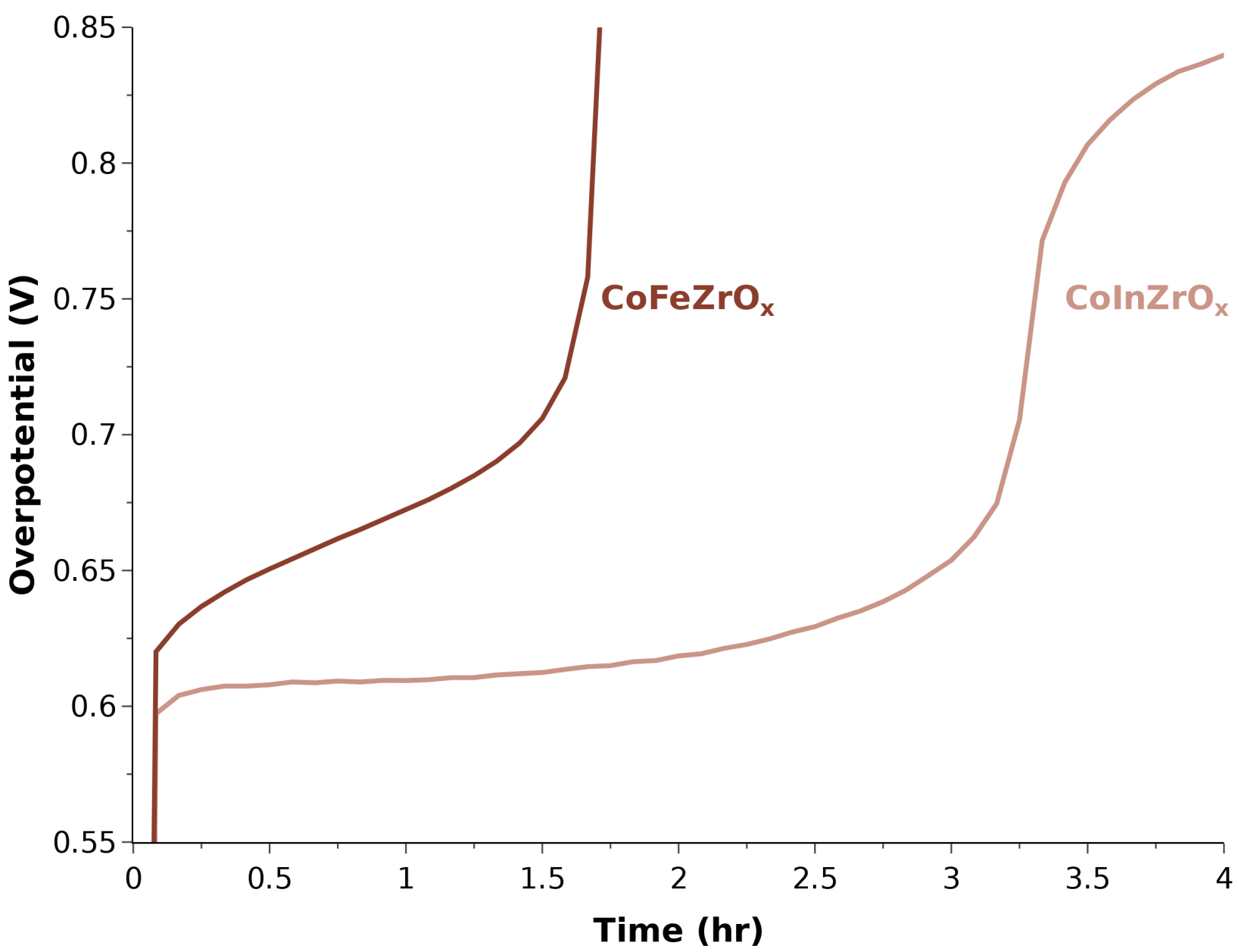

**Fig. S1: Medium-term H-cell testing of selected non-platinum-group-metal (non-PGM) candidates.** Measurements conducted in 1 M $H_2SO_4$, room temperature, 10 mA cm$^{-2}$. CoInZrO$_x$ centered around $Co_{78}In_{12}Zr_{10}O_x$ and CoFeZrO$_x$ centered around $Co_{86}Fe_7Zr_7O_x$.

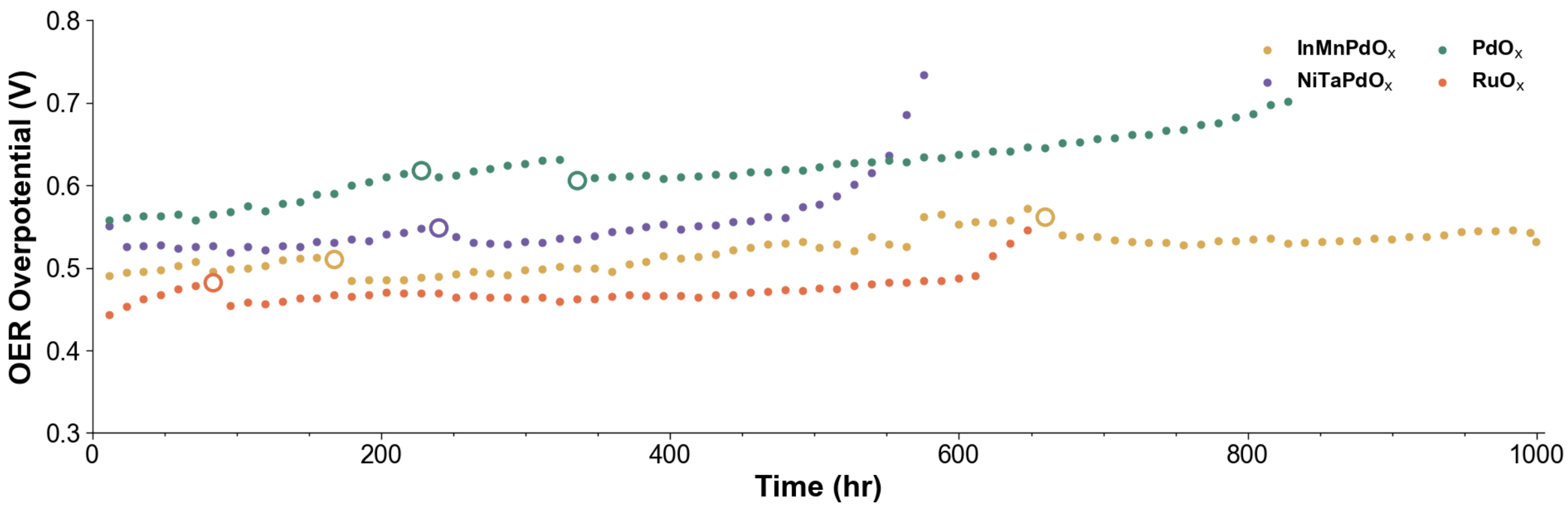

**Fig. S2: Non iR-corrected long-term H-cell testing of Pd-rich front-runners (InMnPdO$_x$ and NiTaPdO$_x$) and benchmark PdO$_x$ and RuO$_x$.** Testing conducted at 10 mA cm$^{-2}$, 1 M $H_2SO_4$, and room temperature with hollow circles indicating interruption events at which the cell was at the open-circuit potential. iR-corrected version shown in Fig. 3E.

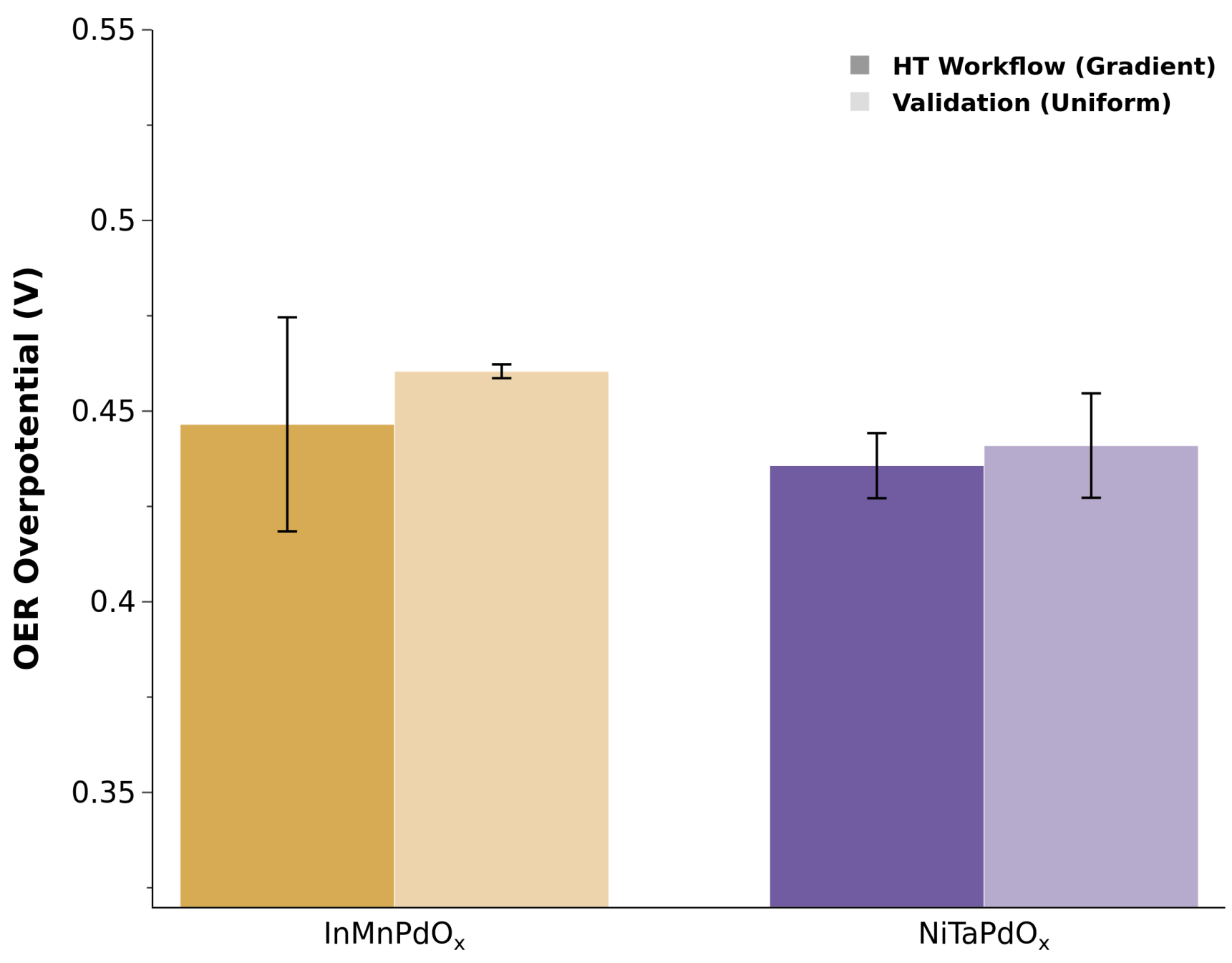

**Fig. S3: SEC reproducibility of Pd front-runners.** Overpotential obtained at the end of SEC screening protocol at 10 mA $cm^{-2}$ for $InMnPdO_x$ and $NiTaPdO_x$ during HT discovery (gradient composition sample) and validation (uniform composition sample) step. Mean and standard deviation are calculated from the 3 tested regions on a 2 × 2 cm sample chip.

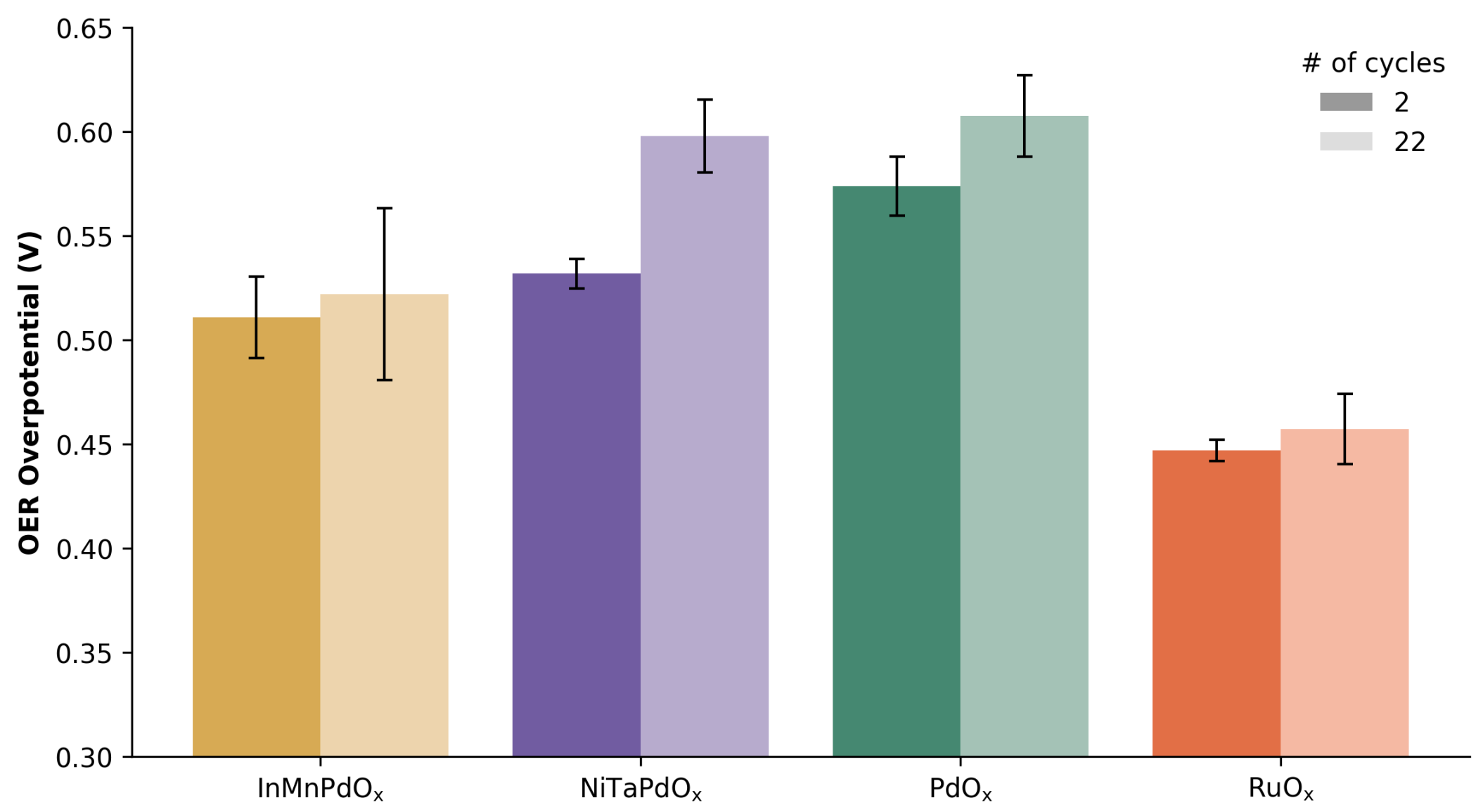

**Fig. S4: Partial duration reproducibility of long-term testing of Pd front-runners and benchmarks ($PdO_x$ and $RuO_x$).** OER overpotential after 2nd (24 h) and 22nd (264 h) 12 h CP cycles. Overpotential value calculated by averaging the last hour of the corresponding 12 h CP. Mean and standard deviation are calculated from triplicate samples (n = 3).

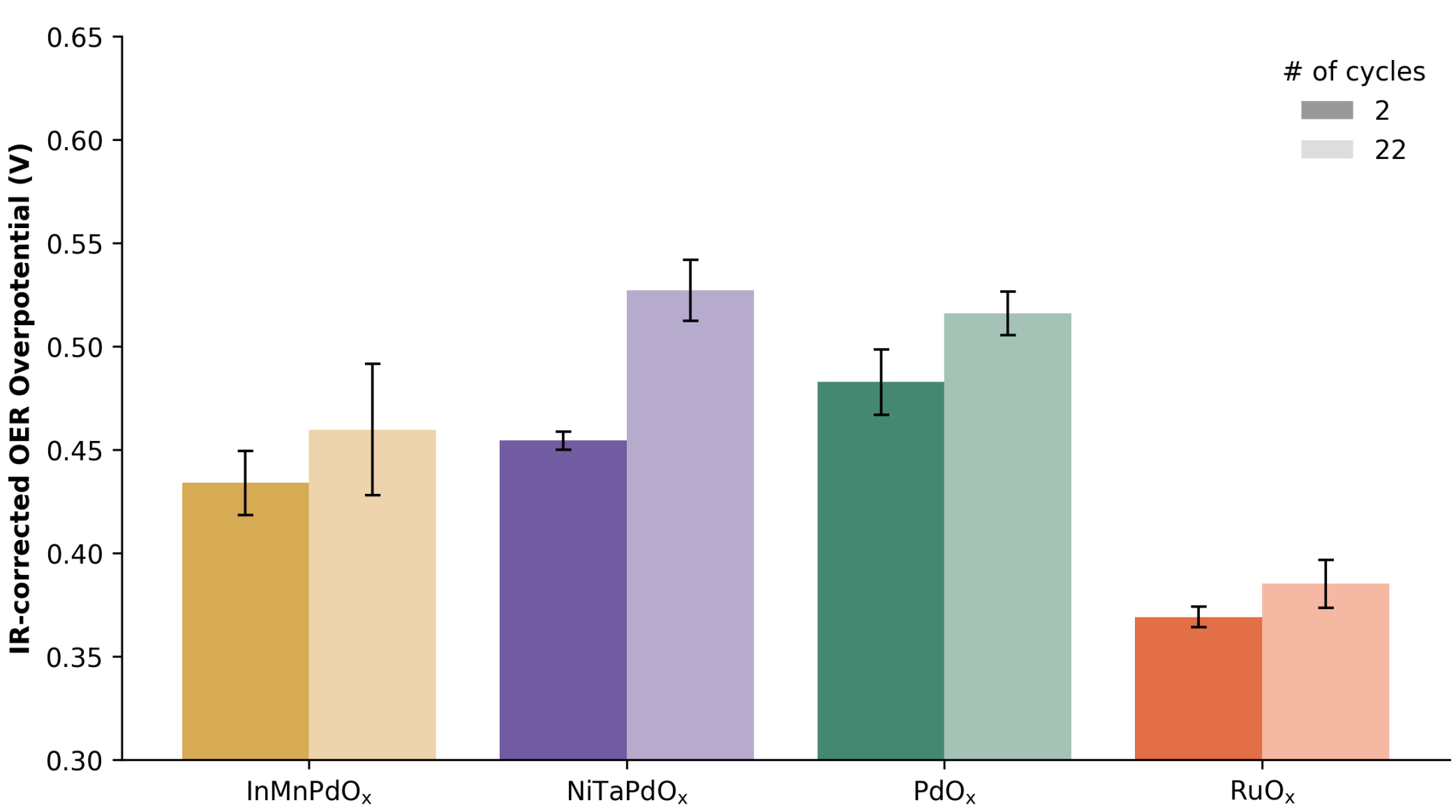

**Fig. S5: iR-corrected partial duration reproducibility of long-term testing of Pd front-runners and benchmarks ($PdO_x$ and $RuO_x$).** iR-corrected OER overpotential after 2$^{nd}$ (24 h) and 22$^{nd}$ (264 h) 12 h CP cycles. Overpotential value calculated by averaging the last hour of the corresponding 12 h CP. Mean and standard deviation are calculated from triplicate samples (n = 3).

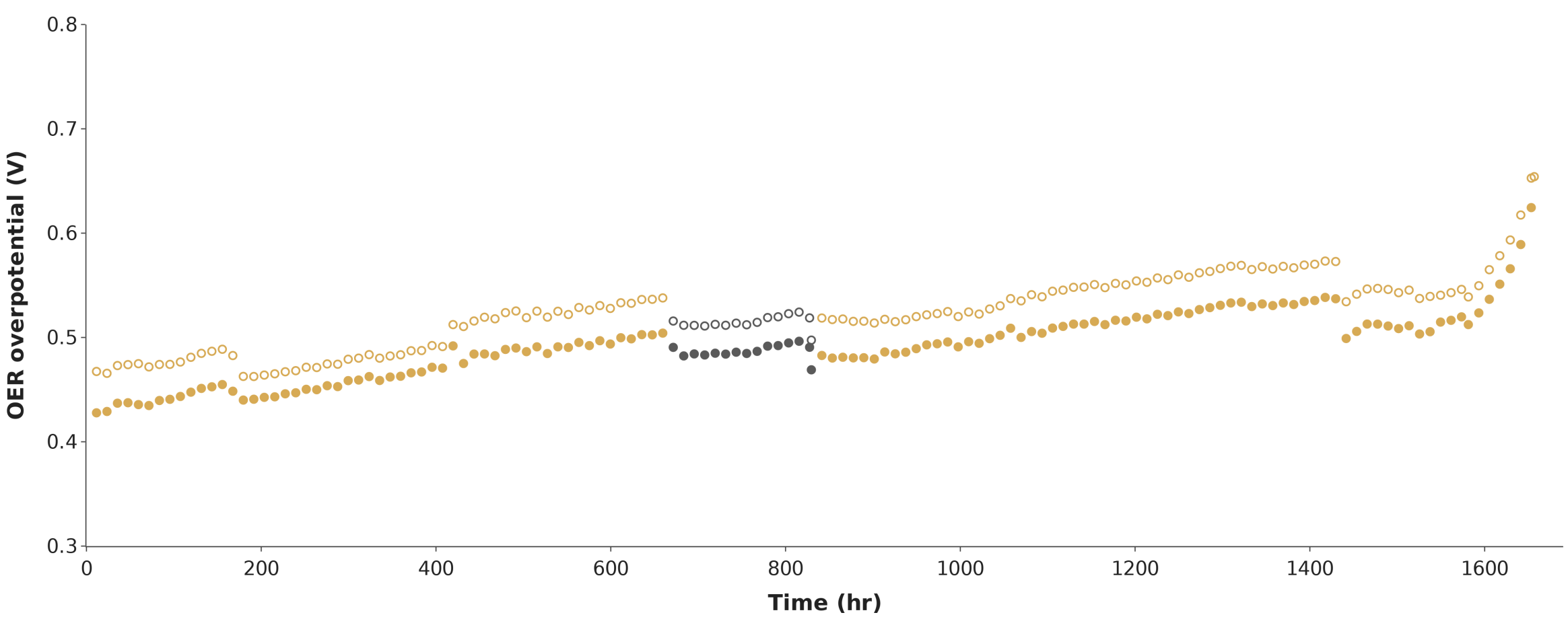

**Fig. S6: Long-term operational stability of InMnPdO$_x$ on platinum-coated Ti PTL.** Measurement conducted in 1 M $H_2SO_4$ at room temperature shown as non iR-corrected (hollow markers) and iR-corrected (filled markers) OER overpotential versus time. Gold markers holding at 10 mA cm$^{-2}$ and grey markers holding at 7.5 mA cm$^{-2}$. The OER overpotential of bare PTL after 12 h at 10 mA cm$^{-2}$ was 0.85 V without iR correction and 0.81 V with iR correction.

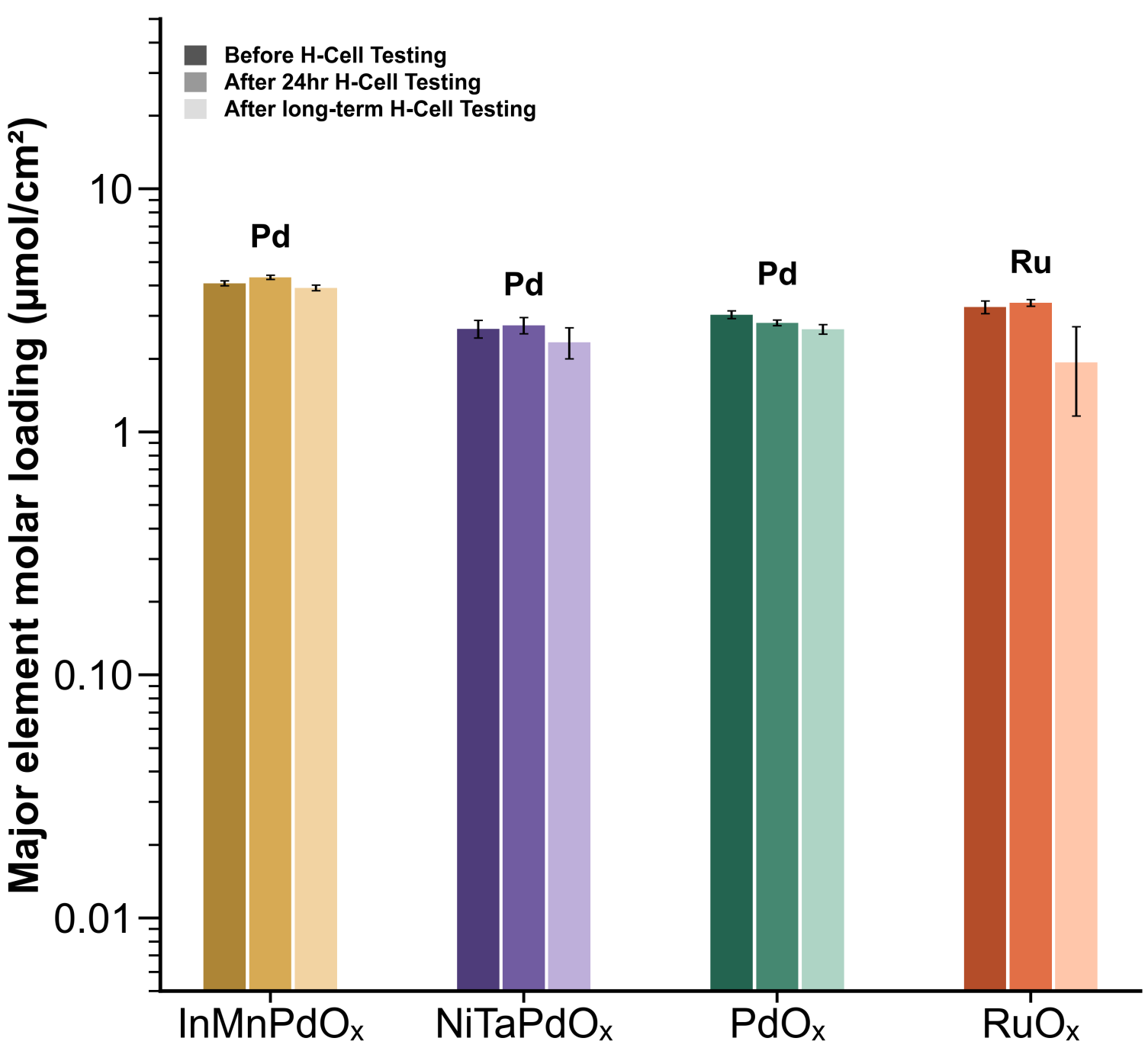

**Fig. S7: XRF molar loadings of major elements (Pd in InMnPdO$_x$, NiTaPdO$_x$, PdO$_x$, and Ru in RuO$_x$) before and after long-term testing at different time stamps.** Error bars represent variation within the 5x5 measurement grid. Note the RuO$_x$ catalyst was partially delaminated from the substrate after long-term testing. XRF measurements for this post-tested sample were only collected on areas with remaining catalyst.

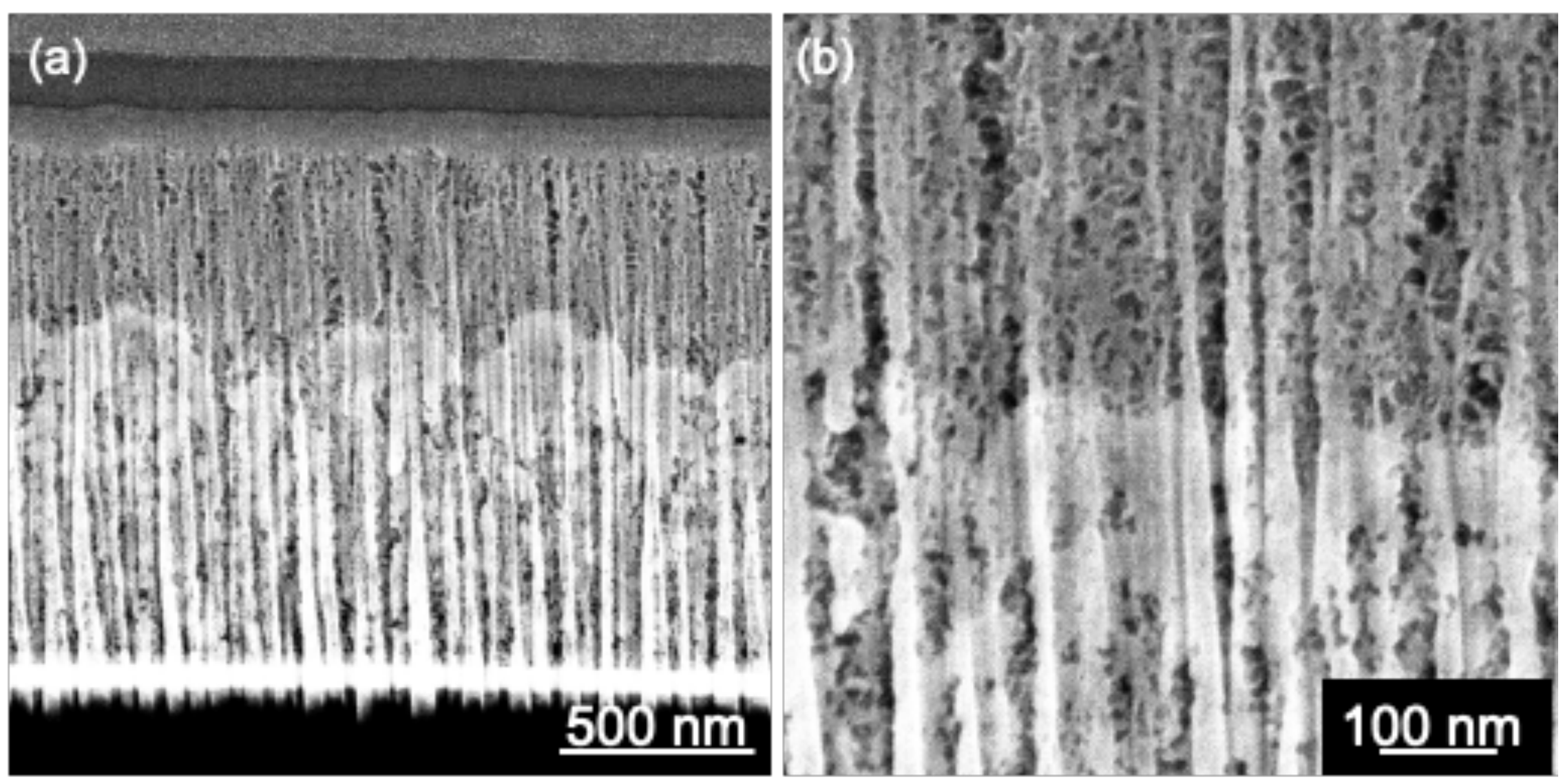

**Fig. S8: SEM image of lift-out lamella post-tested InMnPdO$_x$ sample after long-term testing.** (a) Full view of cross-section. (b) Zoom-in region at the interface of the intermediate layer and the bottom layer.

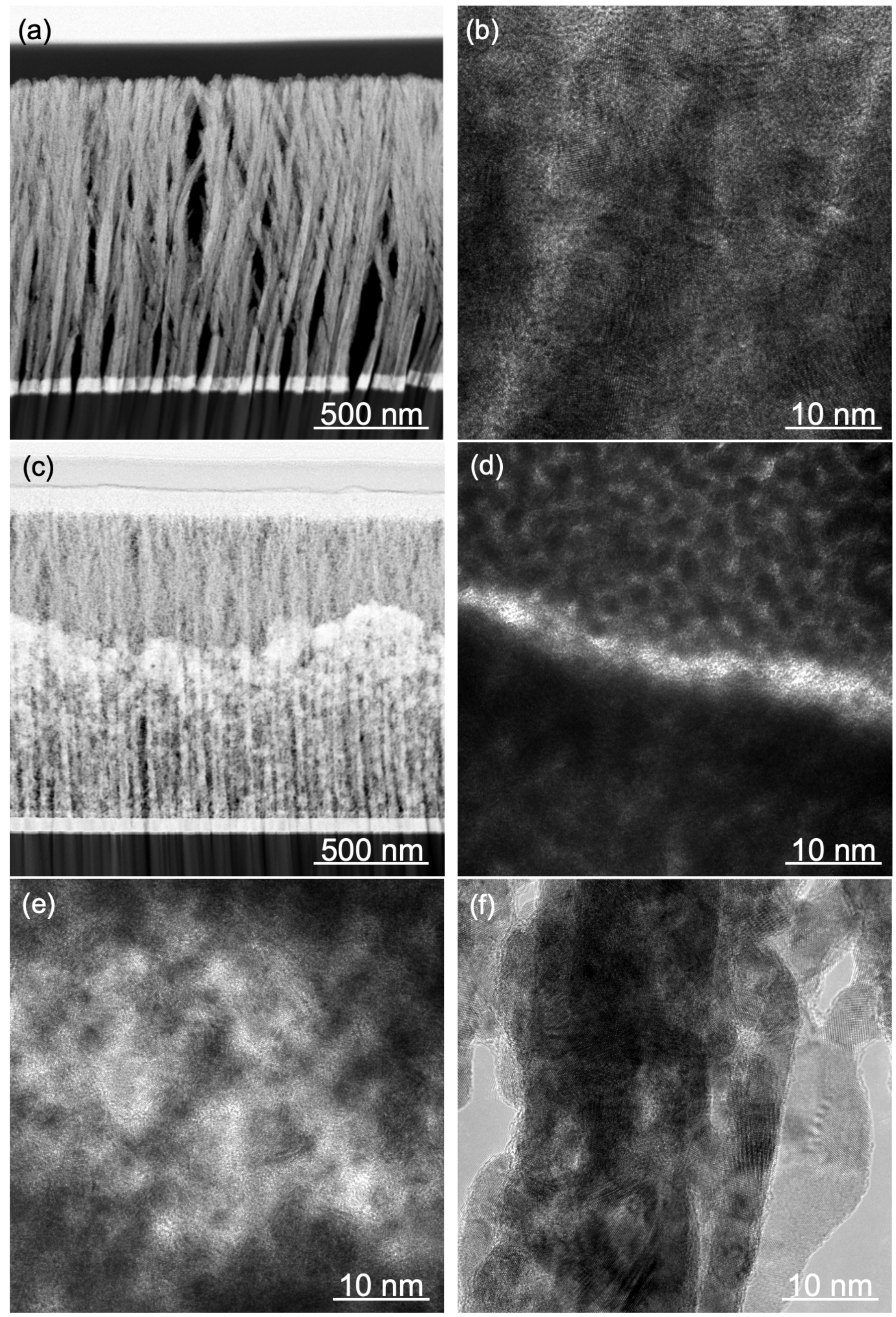

**Fig. S9: STEM and high-resolution TEM (HRTEM) images of as-deposited and post long-term testing of InMnPdO$_x$.** (a) STEM high-angle annular dark-field (HAADF) image of as-deposited sample. (b) HRTEM image near the surface area of as-deposited sample. (c) STEM HAADF image of post-test sample. (d) HRTEM image near surface/Pt-protection layer interface of post-test sample. Note that the surface layer is challenging to thin in focused ion beam (FIB) lift-out without overthinning medium and bottom layers. (e) Post-test InMnPdO$_x$ HRTEM image of the intermediate layer. (f) Post-test InMnPdO$_x$ HRTEM image of the bottom layer.

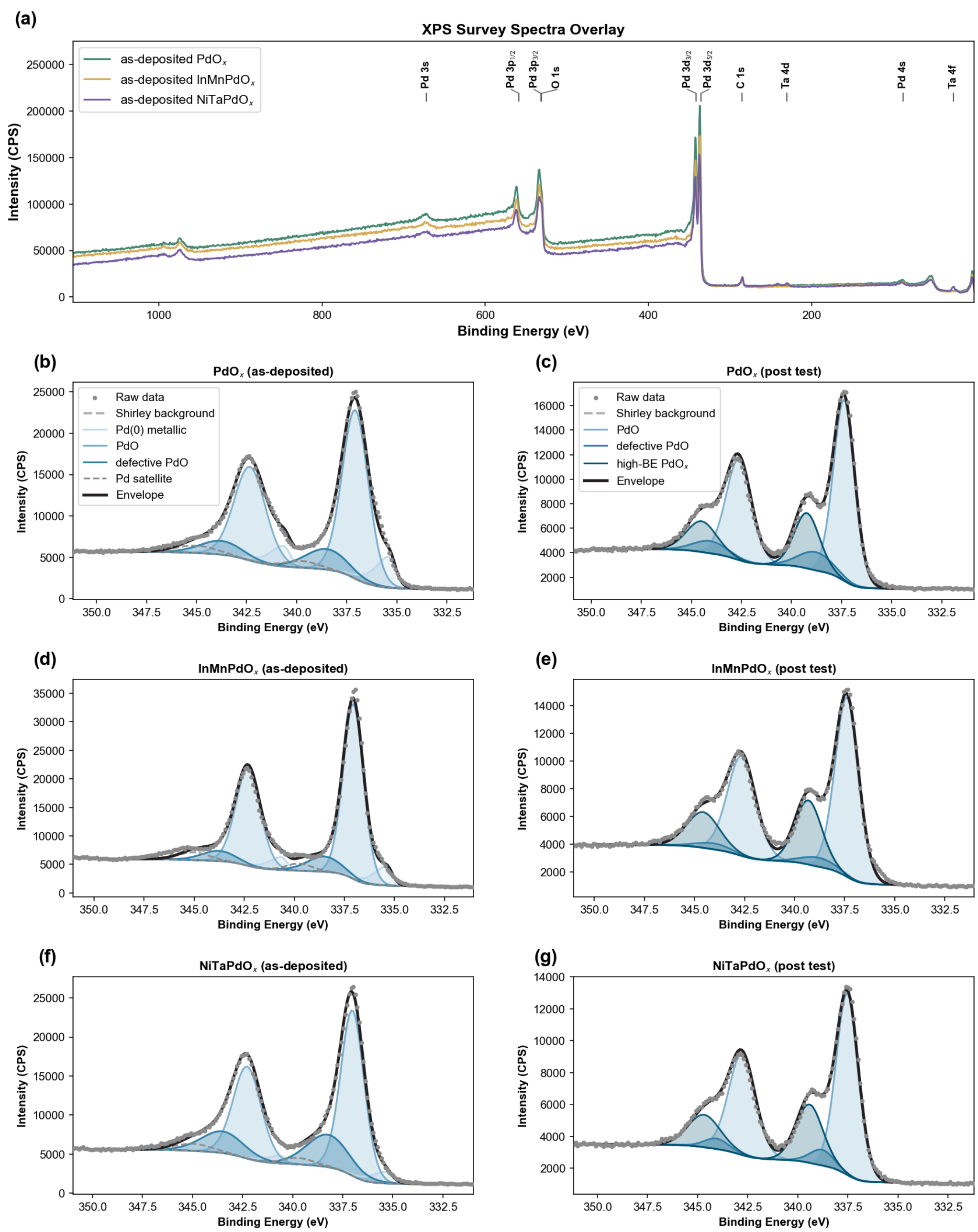

**Fig. S10: XPS spectrum of as-deposited and post-long-term testing for PdO$_x$, InMnPdO$_x$, and NiTaPdO$_x$.** (a) Survey scan of as-deposited samples. High-resolution Pd 3d peak of (b) as-deposited PdO$_x$, (c) post-test PdO$_x$, (d) as-deposited InMnPdO$_x$, (e) post-test InMnPdO$_x$, (f) as-deposited NiTaPdO$_x$, (g) post-test NiTaPdO$_x$. Detailed deconvolution results are listed in Tables S3 and S4.

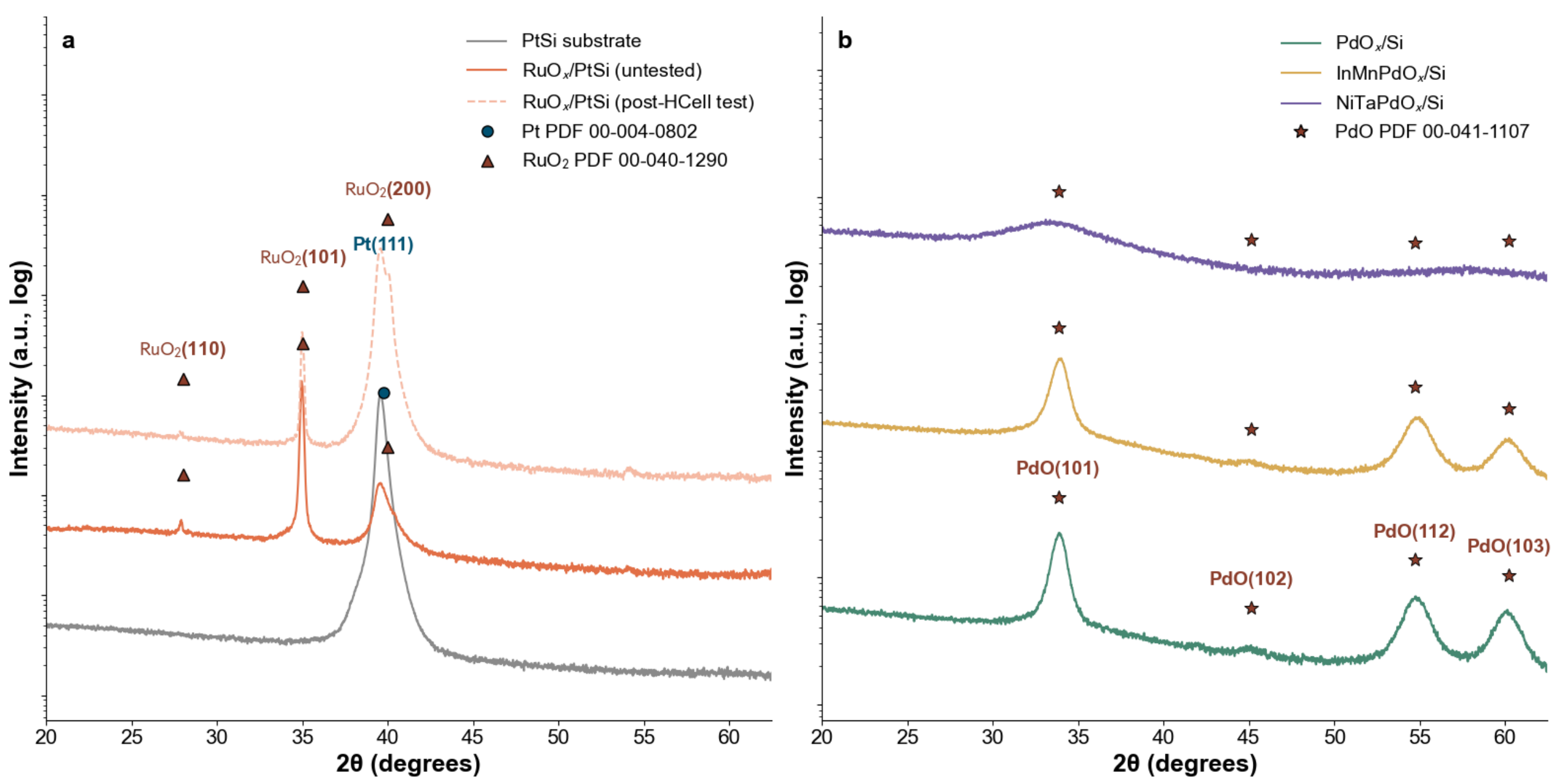

**Fig. S11: XRD spectra of $RuO_x$, $PdO_x$, $InMnPdO_x$, and $NiTaPdO_x$.** (a) $RuO_x$ before and after long-term testing that is shown in Fig. 3E. (b) $PdO_x$, $InMnPdO_x$, and $NiTaPdO_x$ on silicon substrate.

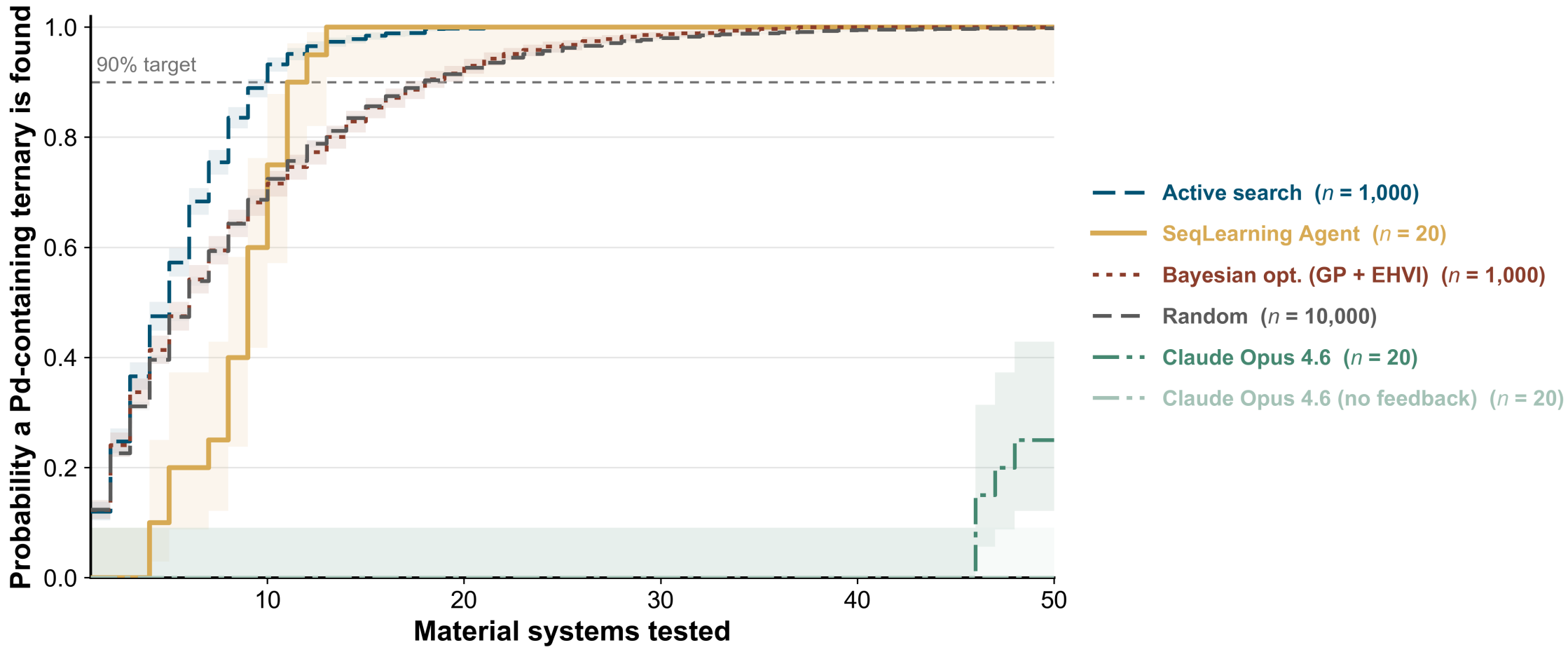

**Fig. S12: Cumulative probability of discovering a Pd-containing catalyst across selection strategies.** Empirical probability that each strategy has selected at least one Pd-containing material system, as a function of material systems tested (50-system budget; dashed horizontal line, 90% target). Each run terminates at its first Pd-containing selection. Curves are empirical fractions over n independent runs per strategy; computationally cheaper strategies were run for more repeats, a purely cost-driven choice. Shaded bands are credible intervals ($5^{th}$–$95^{th}$ percentile) from a Jeffreys Beta (0.5, 0.5) posterior on the per-step discovery probability, and are wider for strategies with smaller n. Where each band crosses the 90% target corresponds to the error-bar endpoints in Fig. 5B. Strategies and interpretation are described in SI section 2.1.

# 4 Supporting Tables

**Table S1: XRD summary of as-deposited InMnPdO$_x$, NiTaPdO$_x$, PdO$_x$, RuO$_x$.** The average crystallite size of as-deposited oxide thin film was estimated by the Scherrer equation (K=0.89). Pd-based oxide thin film matched with tetragonal PdO (PDF 00-041-1107), while RuO$_x$ film matched with $RuO_2$ (PDF 00-040-1290).

| Sample | Dominant oxide peak | Dominant peak position (°) | FWHM (°) | Crystallite size (nm) |
|---|---|---|---|---|
| InMnPdO$_x$ | PdO (101) | 33.92 | 1.16 | 7.08 |
| NiTaPdO$_x$ | N/A | N/A | N/A | N/A |
| PdO$_x$ | PdO (101) | 33.90 | 0.86 | 10.19 |
| RuO$_x$ | $RuO_2$ (101) | 35.01 | 0.14 | 61.34 |

**Table S2: d-spacing calculated from SAED patterns in comparison with reference PdO (00-041-1107).**

| | PdO reference | | As-deposited InMnPdO$_x$ | | Post-test InMnPdO$_x$ (Top) | | Post-test InMnPdO$_x$ (Middle) | | Post-test InMnPdO$_x$ (Bottom) | |
|---|---|---|---|---|---|---|---|---|---|---|
| **Ring #** | **1/d (nm$^{-1}$)** | **d-spacing (Å)** | **1/d (nm$^{-1}$)** | **d-spacing (Å)** | **1/d (nm$^{-1}$)** | **d-spacing (Å)** | **1/d (nm$^{-1}$)** | **d-spacing (Å)** | **1/d (nm$^{-1}$)** | **d-spacing (Å)** |
| 1 | 3.78 (101) | 2.65 | 3.75 | 2.67 | 4.35 | 2.30 | 4.41 | 2.27 | 4.37 | 2.29 |
| 2 | 4.64 (110) | 2.15 | 4.41 | 2.27 | 7.07 | 1.41 | 7.2 | 1.39 | 5.1 | 1.96 |
| 3 | 5.97 (112) | 1.68 | 5.98 | 1.67 | 8.31 | 1.20 | 8.35 | 1.2 | 7.4 | 1.35 |
| 4 | 6.56 (200) | 1.52 | 6.64 | 1.51 | 10.9 | 0.92 | 11 | 0.91 | 8.42 | 1.19 |
| 5 | 7.56 (202) | 1.32 | 7.40 | 1.35 | | | | | 11.2 | 0.89 |
| 6 | | | 8.41 | 1.19 | | | | | | |

**Table S3: XPS Pd 3d peak analysis.** Results for as-deposited and post-tested $PdO_x$, $InMnPdO_x$, and $NiTaPdO_x$ samples after long-term measurement. Peak analysis was conducted using CasaXPS, Shirley background, and energy shift were calibrated by C1s peak at 284.8 eV.

| Sample | | Pd(0) $3d_{3/2}$ | Pd(0) $3d_{5/2}$ | PdO $3d_{3/2}$ | PdO $3d_{5/2}$ | defective PdO $3d_{3/2}$ | defective PdO $3d_{5/2}$ | $PdO_x$ $3d_{3/2}$ | $PdO_x$ $3d_{5/2}$ | Pd Sat. $3d_{3/2}$ | Pd Sat. $3d_{5/2}$ |
|---|---|---|---|---|---|---|---|---|---|---|---|
| As-dep. $PdO_x$ | at.% | 4.40 | 6.62 | 28.05 | 42.17 | 5.54 | 8.32 | | | 2.80 | 2.11 |
| | Raw Area (CPS*eV) | 3545.92 | 5318.87 | 22618.35 | 33927.52 | 4465.32 | 6697.98 | | | 2261.83 | 1696.38 |
| | Position (eV) | 340.66 | 335.4 | 342.33 | 337.07 | 343.76 | 338.50 | | | 345.03 | 339.77 |
| | FWHM | 0.82 | 0.82 | 1.84 | 1.54 | 2.50 | 2.50 | | | 2.34 | 1.95 |
| Post-test $PdO_x$ | at.% | | | 26.25 | 39.45 | 4.36 | 6.55 | 9.48 | 14.25 | | |
| | Raw Area (CPS*eV) | | | 13364.72 | 20047.08 | 2149.38 | 3224.07 | 4832.29 | 7248.43 | | |
| | Position (eV) | | | 342.63 | 337.37 | 344.06 | 338.80 | 344.49 | 339.23 | | |
| | FWHM | | | 1.5 | 1.25 | 2.00 | 2.00 | 1.78 | 1.48 | | |
| As-dep. $InMnPdO_x$ | at.% | 3.26 | 4.90 | 30.22 | 45.42 | 4.36 | 6.55 | | | 3.02 | 2.27 |
| | Raw Area (CPS*eV) | 2871.43 | 4307.14 | 26641.51 | 39962.26 | 3846.68 | 5770.01 | | | 2664.2 | 1998.1 |
| | Position (eV) | 340.66 | 335.40 | 342.31 | 337.05 | 343.76 | 338.50 | | | 345.01 | 339.75 |
| | FWHM | 0.74 | 0.74 | 1.44 | 1.20 | 2.00 | 2.00 | | | 1.80 | 1.50 |
| Post-test $InMnPdO_x$ | at.% | | | 27.00 | 40.58 | 1.94 | 2.91 | 11.02 | 16.56 | | |
| | Raw Area (CPS*eV) | | | 12995.76 | 19493.64 | 933.57 | 1400.36 | 5306.69 | 7960.04 | | |
| | Position (eV) | | | 342.65 | 337.39 | 344.06 | 338.80 | 344.57 | 339.31 | | |
| | FWHM | | | 1.67 | 1.39 | 2.00 | 2.00 | 1.90 | 1.59 | | |
| As-dep. $NiTaPdO_x$ | at.% | 1.79 | 2.69 | 26.79 | 40.27 | 9.50 | 14.28 | | | 2.68 | 2.01 |
| | Raw Area (CPS*eV) | 1406.59 | 2109.89 | 21061.79 | 31592.68 | 7471.61 | 11207.42 | | | 2106.18 | 1579.63 |
| | Position (eV) | 340.65 | 335.39 | 342.28 | 337.02 | 343.5 | 338.24 | | | 344.98 | 339.72 |
| | FWHM | 0.82 | 0.82 | 1.65 | 1.37 | 2.50 | 2.50 | | | 2.16 | 1.80 |
| Post-test $NiTaPdO_x$ | at.% | | | 26.99 | 40.57 | 2.50 | 3.75 | 10.46 | 15.73 | | |
| | Raw Area (CPS*eV) | | | 10843.66 | 16265.49 | 1003.83 | 1505.74 | 4207.71 | 6311.57 | | |
| | Position (eV) | | | 342.78 | 337.52 | 344.06 | 338.8 | 344.68 | 339.42 | | |
| | FWHM | | | 1.58 | 1.32 | 1.42 | 1.42 | 1.91 | 1.59 | | |

**Table S4: Relative distribution of Pd 3d chemical states in $PdO_x$, $InMnPdO_x$, and $NiTaPdO_x$, as-deposited and after long-term testing.** Values are calculated by summing the 3d5/2 and 3d3/2 components for each species from Table S3, normalizing to 100% by excluding satellite peaks. Fitting conditions and the assignment of the defective-PdO and high-binding-energy $PdO_x$ components are given in SI section 1.3.

| Component | As-dep. $PdO_x$ | Post-test $PdO_x$ | As-dep. $InMnPdO_x$ | Post-test $InMnPdO_x$ | As-dep. $NiTaPdO_x$ | Post-test $NiTaPdO_x$ |
|---|---|---|---|---|---|---|
| **Pd(0)** | 11.6% | 0.0% | 8.6% | 0.0% | 4.7% | 0.00% |
| **PdO** | 73.8% | 65.40% | 79.9% | 67.6% | 70.4% | 67.6% |
| **defective PdO** | 14.6% | 10.9% | 11.5% | 4.9% | 24.9% | 6.3% |
| **high binding energy $PdO_x$** | 0.0% | 23.6% | 0.0% | 27.6% | 0.0% | 26.1% |

## 5 Supporting References

1. N. Fairley, *et al.*, Systematic and collaborative approach to problem solving using X-ray photoelectron spectroscopy. *Applied Surface Science Advances* **5**, 100112 (2021).

2. K. S. Kim, A. F. Gossmann, N. Winograd, X-ray photoelectron spectroscopic studies of palladium oxides and the palladium-oxygen electrode. *Anal. Chem.* **46**, 197–200 (1974).

3. L. S. Kibis, A. I. Stadnichenko, S. V. Koscheev, V. I. Zaikovskii, A. I. Boronin, Highly Oxidized Palladium Nanoparticles Comprising $Pd^{4+}$ Species: Spectroscopic and Structural Aspects, Thermal Stability, and Reactivity. *J. Phys. Chem. C* **116**, 19342–19348 (2012).

4. N. Mårtensson, R. Nyholm, Electron spectroscopic determinations of M and N core-hole lifetimes for the elements Nb—Te ( $Z = 41 - 52$ ). *Phys. Rev. B* **24**, 7121–7134 (1981).

5. K. Kan, *et al.*, Accelerated Characterization of Electrode-Electrolyte Equilibration. *ChemCatChem* **16**, e202301300 (2024).

6. D. Guevarra, *et al.*, Orchestrating nimble experiments across interconnected labs. *Digital Discovery* **2**, 1806–1812 (2023).